\documentclass[twocolumn,amsmath,footinbib,amssymb,aps,prl,superscriptaddress]{revtex4-1}

\usepackage{mathtools}
\usepackage{mathrsfs, amsmath, wasysym}
\usepackage[mathscr]{eucal}
\usepackage{xfrac}
\usepackage{graphicx}
\usepackage{dcolumn}
\usepackage{bm}
\usepackage{color}
\usepackage{tabularx}
\usepackage{natbib}
\usepackage{hyperref}
\usepackage[normalem]{ulem}
\usepackage[all]{hypcap}
\usepackage[dvipsnames,usenames,table]{xcolor}

\newcommand{\nocontentsline}[3]{}
\newcommand{\tocless}[2]{\bgroup\let\addcontentsline=\nocontentsline#1{#2}\egroup}



\begin{document}
\title{Nematicity with a twist: rotational symmetry breaking in a moir\'e superlattice}
\author{Rafael M. Fernandes}
\affiliation{School of Physics and Astronomy, University of Minnesota, Minneapolis,
Minnesota 55455, USA}
\author{J\"orn W. F. Venderbos}
\affiliation{Department of Physics, Drexel University, Philadelphia, PA 19104,
USA}
\affiliation{Department of Materials Science \& Engineering, Drexel University,
Philadelphia, PA 19104, USA}
\begin{abstract}
Motivated by recent reports of nematic order in twisted bilayer graphene
(TBG), we investigate the impact of the triangular moir\'e superlattice
degrees of freedom on nematicity. In TBG, the nematic order parameter
is not Ising-like, as it is the case in tetragonal crystals, but has
a 3-state Potts character related to the threefold rotational symmetry
($C_{3z}$) of the moir\'e superlattice. We find that even in the presence
of static strain that explicitly breaks the $C_{3z}$ symmetry, the
system can still undergo a nematic-flop phase transition that spontaneously
breaks in-plane twofold rotations. Moreover, elastic fluctuations,
manifested as acoustic phonons, mediate a nemato-orbital coupling
that ties the orientation of the nematic director to certain soft
directions in momentum space, rendering the Potts-nematic transition
mean-field and first-order. In contrast to the case of rigid crystals,
the Fermi-surface hot-spots associated with these soft directions
are maximally coupled to the low-energy nematic fluctuations in the
case of the moir\'e superlattice. 
\end{abstract}
\date{\today}

\maketitle
\textit{Introduction.} Twisted bilayer graphene (TBG) offers a tantalizing
platform to explore the combined role of effects typically found separately
in strongly correlated materials, topological matter, and two-dimensional
systems. For ``magic'' twist angles, the phase diagram of TBG displays
a rich landscape, showcasing superconductivity, correlated insulating
behavior, ferromagnetism, and anomalous quantum Hall effect \citep{Cao2018a,Cao2018b,Yankowitz1059,Efetov19,Sharpe19,Young19}.
Similar phases are also realized in other twisted compounds \citep{Zhang19,Kim19,Pablo19,FengWang19,FengWang19_2}.
It is believed that this rich physics arises due to the emergence
of isolated---and possibly topologically non-trivial---nearly-flat
bands in the Brillouin zone associated with the moir\'e superlattice~\cite{dosSantos2007,Bistritzer2011,Mele2011,dosSantos2012,Nam2017,Yuan2018,Po2018,Koshino2018,Zou2018,Kang2018,Rademaker2018,Zhang2019a,Zhang2019b,Lian2019,DasSarma2019,Nandkishore2019,Song2019,Kang2019,Tarnopolsky2019}.
This triangular superlattice, with lattice constant of the order of
$10$ nm, is formed by the AA stacking regions, where two carbon atoms
from the two graphene layers sit atop each other {[}Fig. \ref{fig1}(a){]}.
The very small bandwidth, of about $10$ meV, combined with an estimated
Coulomb energy of tens of meV, indicate that correlations play a crucial
role in TBG \citep{STM_Yazdani19}. Indeed, correlated insulating
phases are observed at nearly all commensurate fillings of the moir\'e
unit cell \citep{Cao2018b,Yankowitz1059,Efetov19}, which can host
eight electrons.

Scanning tunneling microscopy \citep{STM_Perge19,STM_Pasupathy19,STM_Andrei19}
and transport measurements \citep{Pablo_nematics} have recently reported
evidence that the three-fold rotational symmetry of the moir\'e superlattice,
denoted by $C_{3z}$, is broken in different regions of the TBG phase
diagram. Moreover, spontaneous $C_{3z}$ symmetry-breaking has been
invoked to explain the observed Landau level degeneracy at charge
neutrality \citep{Senthil19_C3z,Vishwanath19_nematic}. These observations
are suggestive of an electronic nematic phase, i.e. a correlation-driven
lowering of the point group symmetry of a crystal \citep{Fradkin_review,Fernandes_review19}.
Theoretically, a $C_{3z}$ symmetry-breaking phase has been predicted
by some models \citep{Venderbos18,Dodaro2018,Isobe2018,Kozii2019,Chubukov2019}.
Experimentally, however, it is a difficult task to distinguish spontaneous
nematic order from an explicit broken symmetry caused by strain, whose
presence is ubiquitous in TBG \citep{STM_Andrei19,Zeldov19,Guinea19_strain,Pixley2019}.

\begin{figure}
\includegraphics[width=0.9\columnwidth]{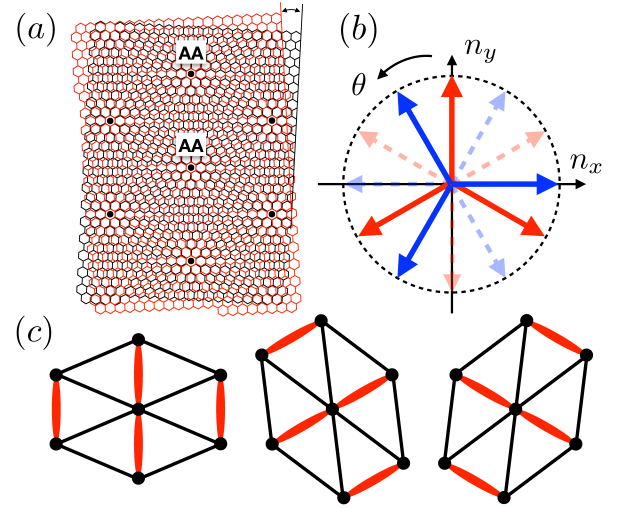} \caption{\label{fig1} (a) Triangular moir\'e superlattice of TBG, formed by
the AA stacking regions (black dots). (b) Allowed directions of the
nematic director $\hat{{\bf n}}=(\cos\theta,\sin\theta)$; blue (red)
corresponds to $\gamma<0$ ($\gamma>0$) in the action Eq. (\ref{S_nem}).
Note that $\mathbf{n}$ and $-\mathbf{n}$ (dashed arrows) are identified.
(c) moir\'e superlattice distortion in the presence of the three symmetry-related
Potts-nematic orders, depicted as a pattern of inequivalent bonds.}
\end{figure}

In this paper, we investigate the interplay between electronic nematic
order and static and fluctuating strain in moir\'e superlattices, applying
our results to TBG. Nematic order in hexagonal (super)lattices, such
as TBG, is fundamentally different from its more well-known counterpart
in tetragonal systems, such as pnictides and cuprates \citep{Fradkin_review,Fernandes14}.
Whereas the latter is described by an Ising order parameter, the former
is described by a two-component order parameter in the 3-state Potts-model
class. As a result, the impact of lattice degrees of freedom is very
different. While static strain completely smears the nematic transition
in tetragonal lattices, it allows the moir\'e superlattice to still
undergo an Ising-like nematic-flop transition, in which in-plane twofold
rotational symmetries are spontaneously broken. Finite-momentum strain fluctuations,
manifested as acoustic phonons, mediate a non-analytic nemato-orbital
coupling in the moir\'e superlattice. The latter makes certain
directions in momentum space -- which are tied to the nematic director's
orientation -- softer than others across the nematic transition. This renders
the 3-state Potts-nematic transition mean-field and first-order, and
also constrains the electronic states that can exchange low-energy
nematic fluctuations to a discrete set of Fermi surface hot-spots.
Because the moir\'e superlattice is not a rigid crystal \citep{Koshino_phonons19,Ochoa19},
the nematic form factor is maximum at these hot-spots. This contrasts
with rigid lattices, where the form factor vanishes at the hot spots,
effectively decoupling the electronic system from low-energy nematic
fluctuations. Thus, the maximum coupling between hot spots and nematic
fluctuations makes moir\'e superlattices promising systems to elucidate
the impact of nematicity on electronic properties.

\textit{Potts-nematic order.} Nematic order is described by a traceless
symmetric tensor, which in two dimensions has two independent components
$\Phi_{1}$ and $\Phi_{2}$ corresponding to the charge quadrupole
moments with $d_{x^{2}-y^{2}}$ and $d_{xy}$ symmetries, respectively.
In systems with tetragonal symmetry, these two $d$-waves have distinct
symmetry and must thus be treated as two independent Ising order parameters.
This is markedly different in hexagonal systems, such as TBG with
point group $D_{6}$~\footnote{The main results presented here also hold if one instead considers
the approach in which TBG is described in terms of the $D_{3}$ point
group.}: the two nematic components belong to a single irreducible representation
of $D_{6}$ and transform as partners under its symmetries, defining
a two-component order parameter $\boldsymbol{\Phi}=(\Phi_{1},\Phi_{2})$.
It is natural to parametrize it as $\boldsymbol{\Phi}=\Phi(\cos2\theta,\sin2\theta)$,
where the angle $\theta$ can be identified with the orientation of
the nematic director $\hat{{\bf n}}=(\cos\theta,\sin\theta)$ (see
Fig.~\ref{fig1}); note that $\boldsymbol{\Phi}\left(\theta\right)=\boldsymbol{\Phi}\left(\theta+\pi\right)$,
as expected.

Although this parametrization might suggest that $\boldsymbol{\Phi}$
is an XY order parameter, the lattice symmetries of TBG introduce
crystal anisotropy effects that pin the nematic director to a discrete
set of high-symmetry directions. Indeed, the Landau-type action $S_{\mathrm{nem}}\left[\boldsymbol{\Phi}\right]$
is (see also  \citep{Hecker18,Venderbos18,Orenstein19,Li_cold_atoms}): 
\begin{equation}
S_{\mathrm{nem}}\left[\boldsymbol{\Phi}\right]=S_{0}\left[\boldsymbol{\Phi}\right]+\frac{\gamma}{6}\int_{x}\left(\Phi_{+}^{3}+\Phi_{-}^{3}\right),\label{S_nem}
\end{equation}
where $x=(\mathbf{r},\tau)$ denotes spatial coordinate $\mathbf{r}$
and imaginary time $\tau$, and $\Phi_{\pm}\equiv\Phi_{1}\pm i\Phi_{2}$.
The first term, $S_{0}\left[\boldsymbol{\Phi}\right]=\frac{1}{2}r_{\Phi}|\boldsymbol{\Phi}|^{2}+\frac{1}{4}u_{\Phi}|\boldsymbol{\Phi}|^{4}$,
is a standard $\Phi^{4}$-action with U(1) symmetry. The cubic term
reflects the crystalline anisotropy of the hexagonal lattice, and
is expressed as $\frac{1}{3}\gamma\Phi^{3}\cos6\theta$, which is
minimized by $\theta=2n\pi/6$ for $\gamma<0$, and $\theta=(2n+1)\pi/6$
for $\gamma>0$. These solutions correspond to sets of threefold degenerate
nematic directors, as shown in Fig.~\ref{fig1}(b) {[}recall that
angles differing by $\pi$ (dashed arrows) must be identified{]},
and manifested as bond orders in real space {[}Fig.~\ref{fig1}(c){]}.
Eq. \eqref{S_nem} is the continuum version of the 3-state Potts-model,
with the $\mathbb{Z}_{3}$ symmetry identified with the out-of-plane
threefold rotation $C_{3z}$.
Below the nematic transition temperature $T_{\text{nem}}$, the sixfold
rotation symmetry $C_{6z}$ is lowered to a twofold symmetry $C_{2z}$,
while the perpendicular twofold rotations $C_{2x}$ and $C_{2y}$
(or their symmetry-related equivalents) are preserved (see inset of
Fig.~\ref{fig2}). Despite the presence of a cubic term in~\eqref{S_nem},
the 3-state Potts transition is continuous in two dimensions \citep{Yu_Potts82}.

\begin{figure}
\includegraphics[width=1\columnwidth]{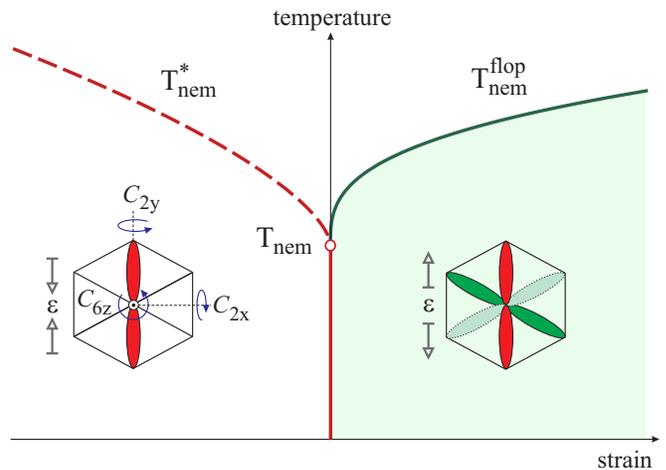} \caption{\label{fig2} Schematic temperature \emph{vs }strain phase diagram
with strain applied along the $y$-axis ($\alpha=\pi/2$) and $\lambda<0$,
$\gamma>0$. For compressive strain ($\varepsilon<0$), because the
director is fixed at $\theta_{0}=\pi/2$, which is a minimum of the
cubic term, no phase transition occurs, and only a crossover temperature
$T_{\mathrm{nem}}^{*}$ survives. For tensile strain ($\varepsilon>0$),
the director is at $\theta_{0}=0$, which is a maximum of the cubic
term, for $T>T_{\mathrm{nem}}^{\mathrm{flop}}$, and at $\pm\bar{\theta}_{0}$
for $T<T_{\mathrm{nem}}^{\mathrm{flop}}$. Thus, $T_{\mathrm{nem}}^{\mathrm{flop}}$
marks an Ising-like nematic-flop transition in which the twofold rotational
symmetries $C_{2x}$ and $C_{2y}$ are spontaneously broken (dark/light
green bonds). The sixfold rotation $C_{6z}$ is explicitly broken
to $C_{2z}$ everywhere for $\varepsilon\protect\neq0$ (red bonds).}
\end{figure}

\textit{Static strain.} As shown in Fig.~\ref{fig1}(c), a lattice
distortion is triggered by nematic order. We include the elastic degrees
of freedom via the strain tensor $\varepsilon_{ij}\equiv\frac{1}{2}(\partial_{i}u_{j}+\partial_{j}u_{i})$
and the rotation tensor $\omega_{ij}\equiv\frac{1}{2}(\partial_{i}u_{j}-\partial_{j}u_{i})$,
where $\mathbf{u}$ is the moir\'e-superlattice displacement vector.
The elasto-nematic action is given by $S_{\mathrm{el-nem}}\left[\boldsymbol{\Phi},\hat{\varepsilon},\hat{\omega}\right]=S_{\mathrm{el}}\left[\hat{\varepsilon},\hat{\omega}\right]+S'\left[\boldsymbol{\Phi},\hat{\varepsilon}\right]$,
where $S_{\mathrm{el}}\left[\hat{\varepsilon},\hat{\omega}\right]$
is the elastic free energy and: 
\begin{equation}
S'\left[\boldsymbol{\Phi},\hat{\varepsilon}\right]=-\lambda\int_{x}\left[\left(\varepsilon_{xx}-\varepsilon_{yy}\right)\Phi_{1}+2\varepsilon_{xy}\Phi_{2}\right].\label{S_el_nem}
\end{equation}
with coupling constant $\lambda$. Consider first the effect of static
strain. For compressive (tensile) uniaxial strain $\varepsilon<0$
($\varepsilon>0$) applied parallel to an arbitrary direction $\hat{\mathbf{d}}$,
the action above becomes $S'=-\lambda\int_{x}\varepsilon\,\Phi\,\cos\left(2\alpha-2\theta\right)$,
where $\cos\alpha=\hat{\mathbf{d}}\cdot\hat{\mathbf{x}}$. At high
temperatures $T\gg T_{\mathrm{nem}}$, where $T_{\mathrm{nem}}$ is
the transition of the unstrained system, we can approximate $S_{\mathrm{nem}}\approx\frac{1}{2}\int_{x}\chi_{\mathrm{nem}}^{-1}\Phi^{2}$.
Thus, strain not only triggers a finite nematic order parameter $\Phi\propto\chi_{\mathrm{nem}}\left|\varepsilon\right|$,
but it also pins the nematic director parallel or perpendicular to
the strain direction, i.e. $\theta_{0}=\alpha$ or $\theta_{0}=\alpha+\pi/2$,
depending on whether $\lambda\varepsilon>0$ or $\lambda\varepsilon<0$,
respectively. To understand what happens as temperature is lowered,
we consider $T\ll T_{\mathrm{nem}}$ and set $\left|\boldsymbol{\Phi}\right|=\Phi_{0}$
as approximately constant. Expanding around the high-temperature director,
$\theta=\theta_{0}+\delta\theta$, gives: 
\begin{equation}
S_{\mathrm{nem}}+S'=\int_{x}\left[a_{\theta_{0}}\left(\delta\theta\right)+b_{\theta_{0}}\left(\delta\theta\right)^{2}\right]\label{static_strain}
\end{equation}
with coefficients $a_{\theta_{0}}=-2\gamma\Phi_{0}^{3}\sin6\theta_{0}$
and $b_{\theta_{0}}=2\Phi_{0}\left(\left|\lambda\varepsilon\right|-3\gamma\Phi_{0}^{2}\cos6\theta_{0}\right)$.
If $\alpha$ (and consequently $\theta_{0}$) does not coincide with
the minima/maxima of the cubic term, i.e. $\alpha\neq n\pi/6$, then
$a_{\theta_{0}}\neq0$. As a result, $\theta$ evolves continuously
from its high-temperature value $\theta_{0}$, and no phase transition
occurs. However, when strain is applied along a high-symmetry direction
($\alpha=n\pi/6$), $a_{\theta_{0}}=0$ and the twofold rotations
$C_{2x}$ and $C_{2y}$ are preserved. These symmetries can nevertheless
be spontaneously broken if $b_{\theta_{0}}<0$. This can only happen
if $\theta_{0}$ coincides with the maxima, but not the minima, of
the cubic term -- in other words, if the strain term $S'$ is minimized
by a director that is maximally penalized by the cubic term of $S_{\mathrm{nem}}$.
In this case, once $\Phi_{0}$ reaches the critical value $\bar{\Phi}_{0}=\sqrt{\left|\frac{\lambda\varepsilon}{3\gamma}\right|}$,
usually at a temperature $T_{\mathrm{nem}}^{\mathrm{flop}}>T_{\mathrm{nem}}$,
the minimum changes from $\theta_{0}$ to $\theta_{0}\pm\bar{\theta}_{0}$,
with $\bar{\theta}_{0}=\frac{1}{2}\arccos\left(\frac{1}{2}\sqrt{1+\frac{3\bar{\Phi}_{0}^{2}}{\Phi_{0}^{2}}}\right)$,
resulting in an Ising-like transition that spontaneously breaks the
$C_{2x}$ and $C_{2y}$ symmetries. Due to its resemblance to the
spin-flop transition, we dub the reorientation of the nematic director
under an external field a nematic-flop transition. Thus, as illustrated
in the phase diagram of Fig. \ref{fig2}, a nematic-driven phase transition
can still occur in a strained triangular lattice \citep{Aharony80,Yu_Potts82},
in contrast to the case of a strained tetragonal lattice, where only
a crossover exists. Therefore, the observation of a spontaneous $C_{2x}$/$C_{2y}$
symmetry-breaking in strained TBG would provide direct evidence for
long-range nematic order.

\begin{figure}
\includegraphics[width=0.9\columnwidth]{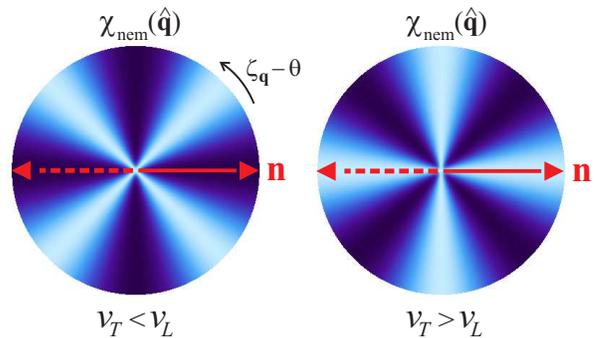}
\caption{\label{fig3} Momentum-directional dependence of the nematic susceptibility
$\chi_{\mathrm{nem}}\left(q\rightarrow0,\,\hat{\mathbf{q}}\right)$
caused by the nemato-orbital coupling, with $\hat{\mathbf{q}}=\left(\cos\zeta_{\mathbf{q}},\,\sin\zeta_{\mathbf{q}}\right)$.
Light blue (dark blue) denotes softer (harder) directions, corresponding
to higher (lower) susceptibility. While for a rigid crystal, $v_{T}<v_{L}$,
the soft direction is rotated by $\pm\pi/4$ with respect to the nematic
director $\mathbf{n}$ (red arrow), for TBG, $v_{T}>v_{L}$, the rotation
is $0,\:\pi/2$. Note that the director can point in any of the directions
$\theta$ of Fig. \ref{fig1}(b).}
\end{figure}

\textit{Fluctuating strain.} Besides static strain, finite-momentum elastic fluctuations
strongly impact the nematic transition \citep{Schmalian16,Paul17,Carvalho19}.
Generally, for a system with $D_{6}$ symmetry, diagonalization of
the harmonic elastic action $S_{\mathrm{el}}\left[\hat{\varepsilon},\hat{\omega}\right]$
leads to two acoustic phonon modes, a transverse ($T$) and a longitudinal
($L$) one with sound velocities $v_{L,T}$: 
\begin{equation}
S_{\mathrm{el}}=\frac{1}{2}\sum_{\mu=L,T}\int_{q}\tilde{u}_{q,\mu}\left(\omega_{n}^{2}+v_{\mu}^{2}\mathbf{q}^{2}\right)\tilde{u}_{-q,\mu},\label{S_phonons}
\end{equation}
where $q=(\mathbf{q},\omega_{n})$, with $\omega_{n}$ the (bosonic)
Matsubara frequency. The displacement field $\mathbf{u}=\sum_{\mu}\tilde{u}_{\mu}\hat{\mathbf{e}}_{\mu}$
has been decomposed into its longitudinal and transverse components
$\tilde{u}_{\mu}$ with $\hat{\mathbf{e}}_{L}=\left(\cos\zeta_{\mathbf{q}},\,\sin\zeta_{\mathbf{q}}\right)$
and $\hat{\mathbf{e}}_{T}=\left(-\sin\zeta_{\mathbf{q}},\,\cos\zeta_{\mathbf{q}}\right)$
and $\zeta_{\mathbf{q}}=\arctan\left(q_{y}/q_{x}\right)$. According
to Refs. \cite{Koshino_phonons19,Ochoa19}, for the dominant acoustic phonons that act on the moir\'e superlattice scale, $\mathbf{u}$ corresponds
to the relative displacement of the two graphene sheets. These and
other phonon modes have been proposed to be linked to superconductivity
in TBG \citep{Lian2019,Wu2018,Frabrizio19,DasSarma_phonons}. Integrating
out the acoustic phonons leads to an additional contribution to the
nematic action, $\delta S_{\mathrm{nem}}=-\frac{1}{2}\sum_{ij}\int_{q}\,\Phi_{i,q}\hat{\Pi}_{ij}\left(q\right)\Phi_{j,-q}$.
In the static limit, $\omega_{n}=0$, we find: 
\begin{equation}
\hat{\Pi}=\frac{\lambda^{2}}{v_{T}^{2}}\left[\hat{\mathbb{I}}-\eta\hat{\mathbb{P}}\right],\quad\hat{\mathbb{P}}=\left(\begin{array}{cc}
\cos^{2}2\zeta_{\mathbf{q}} & \frac{1}{2}\sin4\zeta_{\mathbf{q}}\\
\frac{1}{2}\sin4\zeta_{\mathbf{q}} & \sin^{2}2\zeta_{\mathbf{q}}
\end{array}\right)\label{polarization}
\end{equation}
where $\hat{\mathbb{I}}$ is the identity matrix and $\eta\equiv1-v_{T}^{2}/v_{L}^{2}$.
The first term of $\hat{\Pi}$ gives an overall enhancement of $T_{\mathrm{nem}}$.
The second term couples the two nematic components $\Phi_{1}$ and
$\Phi_{2}$ in a way that depends on the direction, but not on the
magnitude of $\mathbf{q}$. Such a non-analytic term typically appears
when order parameters couple linearly to an elastic mode \citep{Cowley76},
and was previously studied for Ising-nematic order in tetragonal lattices
\citep{Schmalian16,Paul17}. Here, it is manifested as a nemato-orbital
coupling: 
\begin{align}
S_{\mathrm{nem}}^{(\mathrm{eff})}\left[\boldsymbol{\Phi}\right] & =S_{0}\left[\boldsymbol{\Phi}\right]+\frac{\gamma}{6}\int_{x}\left(\Phi_{+}^{3}+\Phi_{-}^{3}\right)\nonumber \\
 & +\frac{\lambda^{2}}{v_{T}^{2}}\left[-\int_{x}\Phi^{2}+\eta\int_{q}\left(\boldsymbol{\Phi}\cdot\hat{\mathbf{D}}\right)^{2}\right]\label{S_nem_eff}
\end{align}
where $\hat{\mathbf{D}}=\left(\cos2\zeta_{\mathbf{q}},\,\sin2\zeta_{\mathbf{q}}\right)=\left(\hat{q}_{x}^{2}-\hat{q}_{y}^{2},2\hat{q}_{x}\hat{q}_{y}\right)$
is the momentum-space (i.e. orbital) quadrupolar form-factor. Recasting
the nemato-orbital coupling term as $\Phi^{2}\cos^{2}\left(2\theta-2\zeta_{\mathbf{q}}\right)$,
we see that it makes only certain directions of momentum space to
become soft, i.e. the static nematic susceptibility $\chi_{\mathrm{nem}}\left(q\rightarrow0,\,\hat{\mathbf{q}}\right)$
is largest near $T_{\mathrm{nem}}$ only along special directions
$\hat{\mathbf{q}}$ (see also \citep{Paul17}). While the cubic term
in Eq. (\ref{S_nem_eff}) forces the director $\hat{\mathbf{n}}$
to point along one of three directions {[}Fig. \ref{fig1}(b){]},
the nemato-orbital coupling makes only two momentum-space directions
$\zeta_{\mathbf{q}}$ soft, namely, the ones that make a relative
angle of $0$ and $\frac{\pi}{2}$ (for $\eta<0$) or $\pm\frac{\pi}{4}$
(for $\eta>0$) with respect to $\hat{\mathbf{n}}$ {[}see Fig. \ref{fig3}{]}.
The reduction of the soft-direction phase-space from continuous to
discrete is known to effectively enhance the dimensionality of the
$\Phi^{4}$-action $S_{0}\left[\boldsymbol{\Phi}\right]$ from $d$
to $d+1$ \citep{Folk76}. Thus, one expects that the nematic transition
in the moir\'e superlattice will be the same as a three-dimensional
3-state Potts-model transition, which is mean-field and first-order
\citep{Yu_Potts82}.

\textit{Electronic degrees of freedom.} If the first-order character of the nematic transition discussed above is weak, nematic fluctuations are still expected to impact the electronic degrees of freedom. For a single-band system with fermionic
operator $c_{\mathbf{k}}$, the electronic-nematic coupling is $S_{\mathrm{elec}}=\int_{\mathbf{k},\mathbf{q}}g\left(\mathbf{k}\right)\Phi_{\mathbf{q}}c_{\mathbf{k}-\mathbf{q}/2}^{\dagger}c_{\mathbf{k}+\mathbf{q}/2}$,
with form factor $g\left(\mathbf{k}\right)=g_{0}\cos\left(2\theta-2\theta_{\mathbf{k}}\right)$,
where $g_{0}$ is a constant and $\theta_{\mathbf{k}}=\arctan(k_{y}/k_{x})$.
The electronic states that exchange low-energy nematic fluctuations
are at the Fermi surface and separated by the small momentum $\mathbf{q}$
of the nematic mode. Since the nematic fluctuations are the softest (albeit non-diverging)
along the special directions $\hat{\mathbf{q}}^{(0)}=\left(\cos\zeta_{\mathbf{q}}^{(0)},\,\sin\zeta_{\mathbf{q}}^{(0)}\right)$
discussed above, the relevant pairs of fermions are located around
the ``hot spots'' $\mathbf{k}_{\mathrm{hs}}$ where the Fermi surface's
tangent is parallel to $\hat{\mathbf{q}}^{(0)}$, i.e. $\hat{\mathbf{q}}^{(0)}\cdot\boldsymbol{\nabla}\xi_{\mathbf{k}_{\mathrm{hs}}}=0$.
The issue is how strong these fermions are coupled to the nematic
fluctuations, i.e. what is the magnitude of $g\left(\mathbf{k}_{\mathrm{hs}}\right)$.
For a circular Fermi surface, the hot spots are located at $\theta_{\mathbf{k}_{\mathrm{hs}}}=\zeta_{\mathbf{q}}^{(0)}+\frac{\pi}{2}$,
and thus $g\left(\mathbf{k}_{\mathrm{hs}}\right)=-g_{0}\cos\left(2\theta-2\zeta_{\mathbf{q}}^{(0)}\right)$.
As we saw above, if $\eta>0$, the soft directions are $\zeta_{\mathbf{q}}^{(0)}=\theta\pm\pi/4$,
yielding $g\left(\mathbf{k}_{\mathrm{hs}}\right)=0$. Thus, in this
case, the hot spots effectively decouple from the softest nematic
fluctuations, similarly to what was obtained for an Ising-nematic
tetragonal lattice \citep{Paul17}. On the other hand, if $\eta<0$,
the soft directions are $\zeta_{\mathbf{q}}^{(0)}=\theta,\,\theta\pm\pi/2$,
implying that $\left|g\left(\mathbf{k}_{\mathrm{hs}}\right)\right|=\left|g_{0}\right|$,
i.e. the hot spots are maximally coupled to the soft nematic fluctuations.
For a generic non-circular Fermi surface respecting $D_{6}$ symmetry
$g\left(\mathbf{k}_{\mathrm{hs}}\right)$ remains maximum for $\eta<0$,
but is expected to be non-zero albeit small for $\eta>0$.

\begin{figure}
\includegraphics[width=1\columnwidth]{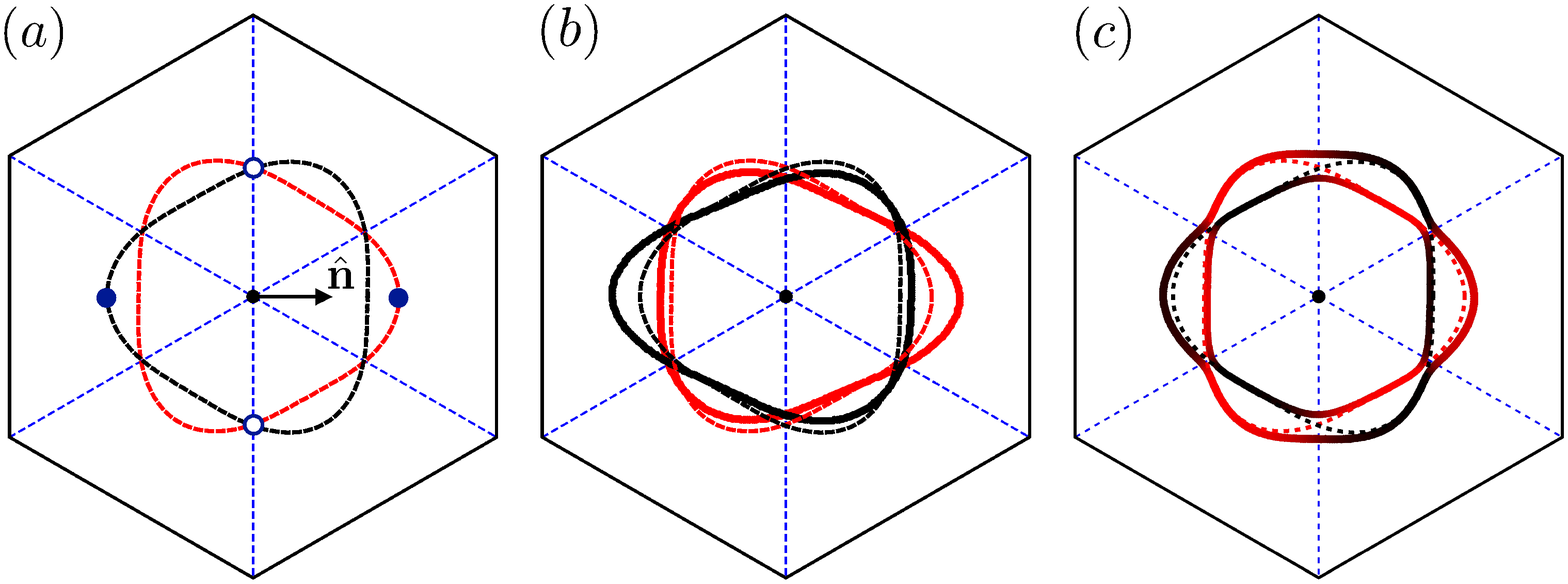} \caption{\label{fig4} (a) Fermi surface of the 6-band model of Ref.~\protect\protect\onlinecite{Po2019};
red and black correspond to the two valleys. The two pairs of hot
spots are marked by open and full symbols. (b) Distortion of the Fermi
surface in the presence of intra-valley Potts-nematic order, with
nematic director $\hat{{\bf n}}$ along the $x$ axis. (c) Same as
panel (b), but for inter-valley nematic order. In (b) and (c), the
undistorted Fermi surface is shown by the dashed lines.}
\end{figure}

The sign of $\eta\equiv1-v_{T}^{2}/v_{L}^{2}$ is determined by the
elastic action $S_{\mathrm{el}}[\hat{\varepsilon},\hat{\omega}]$.
For a rigid crystal, $S_{\mathrm{el}}\left[\hat{\varepsilon},\hat{\omega}\right]=\frac{1}{2}\int_{x}[\left(\partial_{\tau}\mathbf{u}\right)^{2}+C_{ijkl}\varepsilon_{ij}\varepsilon_{kl}]$
depends only on the strain $\hat{\varepsilon}$, since global rotations
do not cost energy. In a triangular lattice, there are only two independent
elastic constants, $C_{11}\equiv C_{xxxx}$ and $C_{12}\equiv C_{xxyy}$,
yielding $v_{L}^{2}=C_{11}$ and $v_{T}^{2}=(C_{11}-C_{12})/2$. Lattice
stability requires $C_{11}>\left|C_{12}\right|$, which makes $\eta>0$,
implying that $g\left(\mathbf{k}_{\mathrm{hs}}\right)$ is small.
However, the moir\'e superlattice is not a rigid crystalline structure
for small twist angles, as lattice relaxation leads to sharp domain
walls separating the regions with AB and BA stacking. Because of this,
arbitrary rotations of the moir\'e superlattice cost energy, and the
elastic free energy acquires an extra term $\delta S_{\mathrm{el}}\left[\hat{\omega}\right]=\frac{1}{2}\int_{x}K\,\omega_{xy}^{2}$
\citep{Ochoa19}. This term contributes only to the transverse velocity
and when $K>2(C_{11}+C_{12})$, $v_{T}$ becomes larger than $v_{L}$
(i.e. $\eta<0$), implying that $\left|g\left(\mathbf{k}_{\mathrm{hs}}\right)\right|$
is maximum. Recent calculations of the acoustic phonon spectrum of
TBG found that this condition is satisfied for small twist angles
\citep{Ochoa19,Koshino_phonons19}, making TBG a rather unique system
in which the Fermi-surface hot spots are maximally coupled to the
nematic fluctuations.

To apply these results to TBG, we use the six-band model of Ref.~\onlinecite{Po2019}.
As shown in Fig.~\ref{fig4}(a), there are two Fermi surfaces associated
with the two valley degrees of freedom, and thus related by a $C_{2z}$
rotation. Because the two pairs of hot spots for a given nematic director
$\theta$ are related by $\pi/2$ rotations, they correspond to different
valley symmetries. Setting $\theta=0$ for concreteness, we find that
the pair of hot spots located at $\theta_{\mathbf{k}_{\mathrm{hs}}}=0,\,\pi$
is associated with intra-valley nematicity {[}Fig.~\ref{fig4}(b){]},
whereas the pair located at $\theta_{\mathbf{k}_{\mathrm{hs}}}=\pm\pi/2$
is associated with inter-valley nematicity {[}Fig.~\ref{fig4}(c){]}
(for details, see Supplementary Material).

We conclude by discussing the possible microscopic mechanisms for
Potts-nematic order in TBG. In weak-coupling approaches, a Pomeranchuk-instability
breaking the $C_{3z}$ rotational-symmetry of the Fermi surface can
be favored by van Hove singularities \citep{Valenzuela08}. In strong-coupling
approaches, where charge degrees of freedom are quenched, a widely
used effective Hamiltonian is described in terms of an SU(4) ``super-spin''
associated with spin and orbital variables \citep{Xu2018a,Venderbos18,Kang2019,Classen19,Kiese19,Natori19}.
Nematicity is then described by an ordering of the orbital variables,
i.e. ordering in the SU(2) orbital sector, which breaks spatial rotational
symmetry. Whether the ground state of the effective SU(4) Hamiltonian
is a nematic phase is an interesting open question. A third possible
mechanism is a nematic phase that is a vestigial order of a primary
electronic ordered state that breaks $C_{3z}$ and some additional
symmetry \citep{Fernandes_review19}, such as $p+p$-wave/$d+d$-wave
superconductivity \citep{Hecker18,Venderbos18,Chubukov2019} or stripe
spin density-waves \citep{Orenstein19}.

\textit{Conclusions. } We showed that the Potts-like
character of the nematic order parameter in triangular moir\'e superlattices
leads to unique nematic behaviors seen neither in tetragonal systems
nor in rigid triangular crystals. Notably, a nematic-flop phase transition
that spontaneously breaks the in-plane twofold rotational symmetries
can still take place even when $C_{3z}$ symmetry-breaking strain
is applied. This makes it possible to unambiguously detect long-range
nematic order in TBG despite the unavoidable presence of residual
strains that break the same symmetry as the nematic order parameter.
Moreover, the emergence of a nemato-orbital coupling mediated by acoustic
phonons affects not only the character of the Potts-nematic transition,
which becomes mean-field and first-order, but also the impact of the
low-energy nematic fluctuations on the electronic properties, which
is maximized due to the non-rigid nature of the moir\'e superlattice. 
\begin{acknowledgments}
We thank A. Chubukov, L. Fu, P. Jarillo-Herrero, J. Kang, H. Ochoa,
H. C. Po, J. Schmalian, T. Senthil, and O. Vafek for fruitful discussions.
R.M.F. was supported by the U. S. Department of Energy, Office of
Science, Basic Energy Sciences, under Award No. DE-SC0020045. 
\end{acknowledgments}

\bibliography{references}

\begin{thebibliography}{70}%
\makeatletter
\providecommand \@ifxundefined [1]{%
 \@ifx{#1\undefined}
}%
\providecommand \@ifnum [1]{%
 \ifnum #1\expandafter \@firstoftwo
 \else \expandafter \@secondoftwo
 \fi
}%
\providecommand \@ifx [1]{%
 \ifx #1\expandafter \@firstoftwo
 \else \expandafter \@secondoftwo
 \fi
}%
\providecommand \natexlab [1]{#1}%
\providecommand \enquote  [1]{``#1''}%
\providecommand \bibnamefont  [1]{#1}%
\providecommand \bibfnamefont [1]{#1}%
\providecommand \citenamefont [1]{#1}%
\providecommand \href@noop [0]{\@secondoftwo}%
\providecommand \href [0]{\begingroup \@sanitize@url \@href}%
\providecommand \@href[1]{\@@startlink{#1}\@@href}%
\providecommand \@@href[1]{\endgroup#1\@@endlink}%
\providecommand \@sanitize@url [0]{\catcode `\\12\catcode `\$12\catcode
  `\&12\catcode `\#12\catcode `\^12\catcode `\_12\catcode `\%12\relax}%
\providecommand \@@startlink[1]{}%
\providecommand \@@endlink[0]{}%
\providecommand \url  [0]{\begingroup\@sanitize@url \@url }%
\providecommand \@url [1]{\endgroup\@href {#1}{\urlprefix }}%
\providecommand \urlprefix  [0]{URL }%
\providecommand \Eprint [0]{\href }%
\providecommand \doibase [0]{http://dx.doi.org/}%
\providecommand \selectlanguage [0]{\@gobble}%
\providecommand \bibinfo  [0]{\@secondoftwo}%
\providecommand \bibfield  [0]{\@secondoftwo}%
\providecommand \translation [1]{[#1]}%
\providecommand \BibitemOpen [0]{}%
\providecommand \bibitemStop [0]{}%
\providecommand \bibitemNoStop [0]{.\EOS\space}%
\providecommand \EOS [0]{\spacefactor3000\relax}%
\providecommand \BibitemShut  [1]{\csname bibitem#1\endcsname}%
\let\auto@bib@innerbib\@empty
\bibitem [{\citenamefont {Cao}\ \emph {et~al.}(2018{\natexlab{a}})\citenamefont
  {Cao}, \citenamefont {Fatemi}, \citenamefont {Fang}, \citenamefont
  {Watanabe}, \citenamefont {Taniguchi}, \citenamefont {Kaxiras},\ and\
  \citenamefont {Jarillo-Herrero}}]{Cao2018a}%
  \BibitemOpen
  \bibfield  {author} {\bibinfo {author} {\bibfnamefont {Y.}~\bibnamefont
  {Cao}}, \bibinfo {author} {\bibfnamefont {V.}~\bibnamefont {Fatemi}},
  \bibinfo {author} {\bibfnamefont {S.}~\bibnamefont {Fang}}, \bibinfo {author}
  {\bibfnamefont {K.}~\bibnamefont {Watanabe}}, \bibinfo {author}
  {\bibfnamefont {T.}~\bibnamefont {Taniguchi}}, \bibinfo {author}
  {\bibfnamefont {E.}~\bibnamefont {Kaxiras}}, \ and\ \bibinfo {author}
  {\bibfnamefont {P.}~\bibnamefont {Jarillo-Herrero}},\ }\href {\doibase
  10.1038/nature26160} {\bibfield  {journal} {\bibinfo  {journal} {Nature}\
  }\textbf {\bibinfo {volume} {556}},\ \bibinfo {pages} {43} (\bibinfo {year}
  {2018}{\natexlab{a}})}\BibitemShut {NoStop}%
\bibitem [{\citenamefont {Cao}\ \emph {et~al.}(2018{\natexlab{b}})\citenamefont
  {Cao}, \citenamefont {Fatemi}, \citenamefont {Demir}, \citenamefont {Fang},
  \citenamefont {Tomarken}, \citenamefont {Luo}, \citenamefont
  {Sanchez-Yamagishi}, \citenamefont {Watanabe}, \citenamefont {Taniguchi},
  \citenamefont {Kaxiras}, \citenamefont {Ashoori},\ and\ \citenamefont
  {Jarillo-Herrero}}]{Cao2018b}%
  \BibitemOpen
  \bibfield  {author} {\bibinfo {author} {\bibfnamefont {Y.}~\bibnamefont
  {Cao}}, \bibinfo {author} {\bibfnamefont {V.}~\bibnamefont {Fatemi}},
  \bibinfo {author} {\bibfnamefont {A.}~\bibnamefont {Demir}}, \bibinfo
  {author} {\bibfnamefont {S.}~\bibnamefont {Fang}}, \bibinfo {author}
  {\bibfnamefont {S.~L.}\ \bibnamefont {Tomarken}}, \bibinfo {author}
  {\bibfnamefont {J.~Y.}\ \bibnamefont {Luo}}, \bibinfo {author} {\bibfnamefont
  {J.~D.}\ \bibnamefont {Sanchez-Yamagishi}}, \bibinfo {author} {\bibfnamefont
  {K.}~\bibnamefont {Watanabe}}, \bibinfo {author} {\bibfnamefont
  {T.}~\bibnamefont {Taniguchi}}, \bibinfo {author} {\bibfnamefont
  {E.}~\bibnamefont {Kaxiras}}, \bibinfo {author} {\bibfnamefont {R.~C.}\
  \bibnamefont {Ashoori}}, \ and\ \bibinfo {author} {\bibfnamefont
  {P.}~\bibnamefont {Jarillo-Herrero}},\ }\href {\doibase 10.1038/nature26154}
  {\bibfield  {journal} {\bibinfo  {journal} {Nature}\ }\textbf {\bibinfo
  {volume} {556}},\ \bibinfo {pages} {80} (\bibinfo {year}
  {2018}{\natexlab{b}})}\BibitemShut {NoStop}%
\bibitem [{\citenamefont {Yankowitz}\ \emph {et~al.}(2019)\citenamefont
  {Yankowitz}, \citenamefont {Chen}, \citenamefont {Polshyn}, \citenamefont
  {Zhang}, \citenamefont {Watanabe}, \citenamefont {Taniguchi}, \citenamefont
  {Graf}, \citenamefont {Young},\ and\ \citenamefont {Dean}}]{Yankowitz1059}%
  \BibitemOpen
  \bibfield  {author} {\bibinfo {author} {\bibfnamefont {M.}~\bibnamefont
  {Yankowitz}}, \bibinfo {author} {\bibfnamefont {S.}~\bibnamefont {Chen}},
  \bibinfo {author} {\bibfnamefont {H.}~\bibnamefont {Polshyn}}, \bibinfo
  {author} {\bibfnamefont {Y.}~\bibnamefont {Zhang}}, \bibinfo {author}
  {\bibfnamefont {K.}~\bibnamefont {Watanabe}}, \bibinfo {author}
  {\bibfnamefont {T.}~\bibnamefont {Taniguchi}}, \bibinfo {author}
  {\bibfnamefont {D.}~\bibnamefont {Graf}}, \bibinfo {author} {\bibfnamefont
  {A.~F.}\ \bibnamefont {Young}}, \ and\ \bibinfo {author} {\bibfnamefont
  {C.~R.}\ \bibnamefont {Dean}},\ }\href {\doibase 10.1126/science.aav1910}
  {\bibfield  {journal} {\bibinfo  {journal} {Science}\ }\textbf {\bibinfo
  {volume} {363}},\ \bibinfo {pages} {1059} (\bibinfo {year}
  {2019})}\BibitemShut {NoStop}%
\bibitem [{\citenamefont {Lu}\ \emph {et~al.}(2019)\citenamefont {Lu},
  \citenamefont {Stepanov}, \citenamefont {Yang}, \citenamefont {Xie},
  \citenamefont {Aamir}, \citenamefont {Das}, \citenamefont {Urgell},
  \citenamefont {Watanabe}, \citenamefont {Taniguchi}, \citenamefont {Zhang},
  \citenamefont {Bachtold}, \citenamefont {MacDonald},\ and\ \citenamefont
  {Efetov}}]{Efetov19}%
  \BibitemOpen
  \bibfield  {author} {\bibinfo {author} {\bibfnamefont {X.}~\bibnamefont
  {Lu}}, \bibinfo {author} {\bibfnamefont {P.}~\bibnamefont {Stepanov}},
  \bibinfo {author} {\bibfnamefont {W.}~\bibnamefont {Yang}}, \bibinfo {author}
  {\bibfnamefont {M.}~\bibnamefont {Xie}}, \bibinfo {author} {\bibfnamefont
  {M.~A.}\ \bibnamefont {Aamir}}, \bibinfo {author} {\bibfnamefont
  {I.}~\bibnamefont {Das}}, \bibinfo {author} {\bibfnamefont {C.}~\bibnamefont
  {Urgell}}, \bibinfo {author} {\bibfnamefont {K.}~\bibnamefont {Watanabe}},
  \bibinfo {author} {\bibfnamefont {T.}~\bibnamefont {Taniguchi}}, \bibinfo
  {author} {\bibfnamefont {G.}~\bibnamefont {Zhang}}, \bibinfo {author}
  {\bibfnamefont {A.}~\bibnamefont {Bachtold}}, \bibinfo {author}
  {\bibfnamefont {A.~H.}\ \bibnamefont {MacDonald}}, \ and\ \bibinfo {author}
  {\bibfnamefont {D.~K.}\ \bibnamefont {Efetov}},\ }\href@noop {} {\bibfield
  {journal} {\bibinfo  {journal} {Nature}\ }\textbf {\bibinfo {volume} {574}},\
  \bibinfo {pages} {653} (\bibinfo {year} {2019})}\BibitemShut {NoStop}%
\bibitem [{\citenamefont {Sharpe}\ \emph {et~al.}(2019)\citenamefont {Sharpe},
  \citenamefont {Fox}, \citenamefont {Barnard}, \citenamefont {Finney},
  \citenamefont {Watanabe}, \citenamefont {Taniguchi}, \citenamefont
  {Kastner},\ and\ \citenamefont {Goldhaber-Gordon}}]{Sharpe19}%
  \BibitemOpen
  \bibfield  {author} {\bibinfo {author} {\bibfnamefont {A.~L.}\ \bibnamefont
  {Sharpe}}, \bibinfo {author} {\bibfnamefont {E.~J.}\ \bibnamefont {Fox}},
  \bibinfo {author} {\bibfnamefont {A.~W.}\ \bibnamefont {Barnard}}, \bibinfo
  {author} {\bibfnamefont {J.}~\bibnamefont {Finney}}, \bibinfo {author}
  {\bibfnamefont {K.}~\bibnamefont {Watanabe}}, \bibinfo {author}
  {\bibfnamefont {T.}~\bibnamefont {Taniguchi}}, \bibinfo {author}
  {\bibfnamefont {M.~A.}\ \bibnamefont {Kastner}}, \ and\ \bibinfo {author}
  {\bibfnamefont {D.}~\bibnamefont {Goldhaber-Gordon}},\ }\href {\doibase
  10.1126/science.aaw3780} {\bibfield  {journal} {\bibinfo  {journal}
  {Science}\ }\textbf {\bibinfo {volume} {365}},\ \bibinfo {pages} {605}
  (\bibinfo {year} {2019})}\BibitemShut {NoStop}%
\bibitem [{\citenamefont {Serlin}\ \emph {et~al.}(2019)\citenamefont {Serlin},
  \citenamefont {Tschirhart}, \citenamefont {Polshyn}, \citenamefont {Zhang},
  \citenamefont {Zhu}, \citenamefont {Watanabe}, \citenamefont {Taniguchi},
  \citenamefont {Balents},\ and\ \citenamefont {Young}}]{Young19}%
  \BibitemOpen
  \bibfield  {author} {\bibinfo {author} {\bibfnamefont {M.}~\bibnamefont
  {Serlin}}, \bibinfo {author} {\bibfnamefont {C.}~\bibnamefont {Tschirhart}},
  \bibinfo {author} {\bibfnamefont {H.}~\bibnamefont {Polshyn}}, \bibinfo
  {author} {\bibfnamefont {Y.}~\bibnamefont {Zhang}}, \bibinfo {author}
  {\bibfnamefont {J.}~\bibnamefont {Zhu}}, \bibinfo {author} {\bibfnamefont
  {K.}~\bibnamefont {Watanabe}}, \bibinfo {author} {\bibfnamefont
  {T.}~\bibnamefont {Taniguchi}}, \bibinfo {author} {\bibfnamefont
  {L.}~\bibnamefont {Balents}}, \ and\ \bibinfo {author} {\bibfnamefont
  {A.}~\bibnamefont {Young}},\ }\href@noop {} {\bibfield  {journal} {\bibinfo
  {journal} {arXiv:1907.00261}\ } (\bibinfo {year} {2019})}\BibitemShut
  {NoStop}%
\bibitem [{\citenamefont {Shen}\ \emph {et~al.}(2019)\citenamefont {Shen},
  \citenamefont {Li}, \citenamefont {Wang}, \citenamefont {Zhao}, \citenamefont
  {Tang}, \citenamefont {Liu}, \citenamefont {Tian}, \citenamefont {Chu},
  \citenamefont {Watanabe}, \citenamefont {Taniguchi}, \citenamefont {Yang},
  \citenamefont {Meng}, \citenamefont {Shi},\ and\ \citenamefont
  {Zhang}}]{Zhang19}%
  \BibitemOpen
  \bibfield  {author} {\bibinfo {author} {\bibfnamefont {C.}~\bibnamefont
  {Shen}}, \bibinfo {author} {\bibfnamefont {N.}~\bibnamefont {Li}}, \bibinfo
  {author} {\bibfnamefont {S.}~\bibnamefont {Wang}}, \bibinfo {author}
  {\bibfnamefont {Y.}~\bibnamefont {Zhao}}, \bibinfo {author} {\bibfnamefont
  {J.}~\bibnamefont {Tang}}, \bibinfo {author} {\bibfnamefont {J.}~\bibnamefont
  {Liu}}, \bibinfo {author} {\bibfnamefont {J.}~\bibnamefont {Tian}}, \bibinfo
  {author} {\bibfnamefont {Y.}~\bibnamefont {Chu}}, \bibinfo {author}
  {\bibfnamefont {K.}~\bibnamefont {Watanabe}}, \bibinfo {author}
  {\bibfnamefont {T.}~\bibnamefont {Taniguchi}}, \bibinfo {author}
  {\bibfnamefont {R.}~\bibnamefont {Yang}}, \bibinfo {author} {\bibfnamefont
  {Z.~Y.}\ \bibnamefont {Meng}}, \bibinfo {author} {\bibfnamefont
  {D.}~\bibnamefont {Shi}}, \ and\ \bibinfo {author} {\bibfnamefont
  {G.}~\bibnamefont {Zhang}},\ }\href@noop {} {\bibfield  {journal} {\bibinfo
  {journal} {arXiv:1903.06952}\ } (\bibinfo {year} {2019})}\BibitemShut
  {NoStop}%
\bibitem [{\citenamefont {Liu}\ \emph {et~al.}(2019{\natexlab{a}})\citenamefont
  {Liu}, \citenamefont {Hao}, \citenamefont {Khalaf}, \citenamefont {Lee},
  \citenamefont {Watanabe}, \citenamefont {Taniguchi}, \citenamefont
  {Vishwanath},\ and\ \citenamefont {Kim}}]{Kim19}%
  \BibitemOpen
  \bibfield  {author} {\bibinfo {author} {\bibfnamefont {X.}~\bibnamefont
  {Liu}}, \bibinfo {author} {\bibfnamefont {Z.}~\bibnamefont {Hao}}, \bibinfo
  {author} {\bibfnamefont {E.}~\bibnamefont {Khalaf}}, \bibinfo {author}
  {\bibfnamefont {J.~Y.}\ \bibnamefont {Lee}}, \bibinfo {author} {\bibfnamefont
  {K.}~\bibnamefont {Watanabe}}, \bibinfo {author} {\bibfnamefont
  {T.}~\bibnamefont {Taniguchi}}, \bibinfo {author} {\bibfnamefont
  {A.}~\bibnamefont {Vishwanath}}, \ and\ \bibinfo {author} {\bibfnamefont
  {P.}~\bibnamefont {Kim}},\ }\href@noop {} {\bibfield  {journal} {\bibinfo
  {journal} {arXiv:1903.08130}\ } (\bibinfo {year}
  {2019}{\natexlab{a}})}\BibitemShut {NoStop}%
\bibitem [{\citenamefont {Cao}\ \emph {et~al.}(2019)\citenamefont {Cao},
  \citenamefont {Rodan-Legrain}, \citenamefont {Rubies-Bigord{\`a}},
  \citenamefont {Park}, \citenamefont {Watanabe}, \citenamefont {Taniguchi},\
  and\ \citenamefont {Jarillo-Herrero}}]{Pablo19}%
  \BibitemOpen
  \bibfield  {author} {\bibinfo {author} {\bibfnamefont {Y.}~\bibnamefont
  {Cao}}, \bibinfo {author} {\bibfnamefont {D.}~\bibnamefont {Rodan-Legrain}},
  \bibinfo {author} {\bibfnamefont {O.}~\bibnamefont {Rubies-Bigord{\`a}}},
  \bibinfo {author} {\bibfnamefont {J.~M.}\ \bibnamefont {Park}}, \bibinfo
  {author} {\bibfnamefont {K.}~\bibnamefont {Watanabe}}, \bibinfo {author}
  {\bibfnamefont {T.}~\bibnamefont {Taniguchi}}, \ and\ \bibinfo {author}
  {\bibfnamefont {P.}~\bibnamefont {Jarillo-Herrero}},\ }\href@noop {}
  {\bibfield  {journal} {\bibinfo  {journal} {arXiv:1903.08596}\ } (\bibinfo
  {year} {2019})}\BibitemShut {NoStop}%
\bibitem [{\citenamefont {Chen}\ \emph
  {et~al.}(2019{\natexlab{a}})\citenamefont {Chen}, \citenamefont {Jiang},
  \citenamefont {Wu}, \citenamefont {Lyu}, \citenamefont {Li}, \citenamefont
  {Chittari}, \citenamefont {Watanabe}, \citenamefont {Taniguchi},
  \citenamefont {Shi}, \citenamefont {Jung}, \citenamefont {Zhang},\ and\
  \citenamefont {Wang}}]{FengWang19}%
  \BibitemOpen
  \bibfield  {author} {\bibinfo {author} {\bibfnamefont {G.}~\bibnamefont
  {Chen}}, \bibinfo {author} {\bibfnamefont {L.}~\bibnamefont {Jiang}},
  \bibinfo {author} {\bibfnamefont {S.}~\bibnamefont {Wu}}, \bibinfo {author}
  {\bibfnamefont {B.}~\bibnamefont {Lyu}}, \bibinfo {author} {\bibfnamefont
  {H.}~\bibnamefont {Li}}, \bibinfo {author} {\bibfnamefont {B.~L.}\
  \bibnamefont {Chittari}}, \bibinfo {author} {\bibfnamefont {K.}~\bibnamefont
  {Watanabe}}, \bibinfo {author} {\bibfnamefont {T.}~\bibnamefont {Taniguchi}},
  \bibinfo {author} {\bibfnamefont {Z.}~\bibnamefont {Shi}}, \bibinfo {author}
  {\bibfnamefont {J.}~\bibnamefont {Jung}}, \bibinfo {author} {\bibfnamefont
  {Y.}~\bibnamefont {Zhang}}, \ and\ \bibinfo {author} {\bibfnamefont
  {F.}~\bibnamefont {Wang}},\ }\href@noop {} {\bibfield  {journal} {\bibinfo
  {journal} {Nature Physics}\ }\textbf {\bibinfo {volume} {15}},\ \bibinfo
  {pages} {237} (\bibinfo {year} {2019}{\natexlab{a}})}\BibitemShut {NoStop}%
\bibitem [{\citenamefont {Chen}\ \emph
  {et~al.}(2019{\natexlab{b}})\citenamefont {Chen}, \citenamefont {Sharpe},
  \citenamefont {Gallagher}, \citenamefont {Rosen}, \citenamefont {Fox},
  \citenamefont {Jiang}, \citenamefont {Lyu}, \citenamefont {Li}, \citenamefont
  {Watanabe}, \citenamefont {Taniguchi}, \citenamefont {Jung}, \citenamefont
  {Shi}, \citenamefont {Goldhaber-Gordon}, \citenamefont {Zhang},\ and\
  \citenamefont {Wang}}]{FengWang19_2}%
  \BibitemOpen
  \bibfield  {author} {\bibinfo {author} {\bibfnamefont {G.}~\bibnamefont
  {Chen}}, \bibinfo {author} {\bibfnamefont {A.~L.}\ \bibnamefont {Sharpe}},
  \bibinfo {author} {\bibfnamefont {P.}~\bibnamefont {Gallagher}}, \bibinfo
  {author} {\bibfnamefont {I.~T.}\ \bibnamefont {Rosen}}, \bibinfo {author}
  {\bibfnamefont {E.~J.}\ \bibnamefont {Fox}}, \bibinfo {author} {\bibfnamefont
  {L.}~\bibnamefont {Jiang}}, \bibinfo {author} {\bibfnamefont
  {B.}~\bibnamefont {Lyu}}, \bibinfo {author} {\bibfnamefont {H.}~\bibnamefont
  {Li}}, \bibinfo {author} {\bibfnamefont {K.}~\bibnamefont {Watanabe}},
  \bibinfo {author} {\bibfnamefont {T.}~\bibnamefont {Taniguchi}}, \bibinfo
  {author} {\bibfnamefont {J.}~\bibnamefont {Jung}}, \bibinfo {author}
  {\bibfnamefont {Z.}~\bibnamefont {Shi}}, \bibinfo {author} {\bibfnamefont
  {D.}~\bibnamefont {Goldhaber-Gordon}}, \bibinfo {author} {\bibfnamefont
  {Y.}~\bibnamefont {Zhang}}, \ and\ \bibinfo {author} {\bibfnamefont
  {F.}~\bibnamefont {Wang}},\ }\href@noop {} {\bibfield  {journal} {\bibinfo
  {journal} {Nature}\ }\textbf {\bibinfo {volume} {10}} (\bibinfo {year}
  {2019}{\natexlab{b}})}\BibitemShut {NoStop}%
\bibitem [{\citenamefont {dos Santos}\ \emph {et~al.}(2007)\citenamefont {dos
  Santos}, \citenamefont {Peres},\ and\ \citenamefont {Neto}}]{dosSantos2007}%
  \BibitemOpen
  \bibfield  {author} {\bibinfo {author} {\bibfnamefont {J.~M. B.~L.}\
  \bibnamefont {dos Santos}}, \bibinfo {author} {\bibfnamefont {N.~M.~R.}\
  \bibnamefont {Peres}}, \ and\ \bibinfo {author} {\bibfnamefont {A.~H.~C.}\
  \bibnamefont {Neto}},\ }\href {\doibase 10.1103/PhysRevLett.99.256802}
  {\bibfield  {journal} {\bibinfo  {journal} {Phys. Rev. Lett.}\ }\textbf
  {\bibinfo {volume} {99}},\ \bibinfo {pages} {256802} (\bibinfo {year}
  {2007})}\BibitemShut {NoStop}%
\bibitem [{\citenamefont {Bistritzer}\ and\ \citenamefont
  {MacDonald}(2011)}]{Bistritzer2011}%
  \BibitemOpen
  \bibfield  {author} {\bibinfo {author} {\bibfnamefont {R.}~\bibnamefont
  {Bistritzer}}\ and\ \bibinfo {author} {\bibfnamefont {A.~H.}\ \bibnamefont
  {MacDonald}},\ }\href {\doibase 10.1073/pnas.1108174108} {\bibfield
  {journal} {\bibinfo  {journal} {Proceedings of the National Academy of
  Sciences}\ }\textbf {\bibinfo {volume} {108}},\ \bibinfo {pages} {12233}
  (\bibinfo {year} {2011})}\BibitemShut {NoStop}%
\bibitem [{\citenamefont {Mele}(2011)}]{Mele2011}%
  \BibitemOpen
  \bibfield  {author} {\bibinfo {author} {\bibfnamefont {E.~J.}\ \bibnamefont
  {Mele}},\ }\href {\doibase 10.1103/PhysRevB.84.235439} {\bibfield  {journal}
  {\bibinfo  {journal} {Phys. Rev. B}\ }\textbf {\bibinfo {volume} {84}},\
  \bibinfo {pages} {235439} (\bibinfo {year} {2011})}\BibitemShut {NoStop}%
\bibitem [{\citenamefont {dos Santos}\ \emph {et~al.}(2012)\citenamefont {dos
  Santos}, \citenamefont {Peres},\ and\ \citenamefont {Neto}}]{dosSantos2012}%
  \BibitemOpen
  \bibfield  {author} {\bibinfo {author} {\bibfnamefont {J.~M. B.~L.}\
  \bibnamefont {dos Santos}}, \bibinfo {author} {\bibfnamefont {N.~M.~R.}\
  \bibnamefont {Peres}}, \ and\ \bibinfo {author} {\bibfnamefont {A.~H.~C.}\
  \bibnamefont {Neto}},\ }\href {\doibase 10.1103/PhysRevB.86.155449}
  {\bibfield  {journal} {\bibinfo  {journal} {Phys. Rev. B}\ }\textbf {\bibinfo
  {volume} {86}},\ \bibinfo {pages} {155449} (\bibinfo {year}
  {2012})}\BibitemShut {NoStop}%
\bibitem [{\citenamefont {Nam}\ and\ \citenamefont {Koshino}(2017)}]{Nam2017}%
  \BibitemOpen
  \bibfield  {author} {\bibinfo {author} {\bibfnamefont {N.~N.~T.}\
  \bibnamefont {Nam}}\ and\ \bibinfo {author} {\bibfnamefont {M.}~\bibnamefont
  {Koshino}},\ }\href {\doibase 10.1103/PhysRevB.96.075311} {\bibfield
  {journal} {\bibinfo  {journal} {Phys. Rev. B}\ }\textbf {\bibinfo {volume}
  {96}},\ \bibinfo {pages} {075311} (\bibinfo {year} {2017})}\BibitemShut
  {NoStop}%
\bibitem [{\citenamefont {Yuan}\ and\ \citenamefont {Fu}(2018)}]{Yuan2018}%
  \BibitemOpen
  \bibfield  {author} {\bibinfo {author} {\bibfnamefont {N.~F.~Q.}\
  \bibnamefont {Yuan}}\ and\ \bibinfo {author} {\bibfnamefont {L.}~\bibnamefont
  {Fu}},\ }\href {\doibase 10.1103/PhysRevB.98.045103} {\bibfield  {journal}
  {\bibinfo  {journal} {Phys. Rev. B}\ }\textbf {\bibinfo {volume} {98}},\
  \bibinfo {pages} {045103} (\bibinfo {year} {2018})}\BibitemShut {NoStop}%
\bibitem [{\citenamefont {Po}\ \emph {et~al.}(2018)\citenamefont {Po},
  \citenamefont {Zou}, \citenamefont {Vishwanath},\ and\ \citenamefont
  {Senthil}}]{Po2018}%
  \BibitemOpen
  \bibfield  {author} {\bibinfo {author} {\bibfnamefont {H.~C.}\ \bibnamefont
  {Po}}, \bibinfo {author} {\bibfnamefont {L.}~\bibnamefont {Zou}}, \bibinfo
  {author} {\bibfnamefont {A.}~\bibnamefont {Vishwanath}}, \ and\ \bibinfo
  {author} {\bibfnamefont {T.}~\bibnamefont {Senthil}},\ }\href {\doibase
  10.1103/PhysRevX.8.031089} {\bibfield  {journal} {\bibinfo  {journal} {Phys.
  Rev. X}\ }\textbf {\bibinfo {volume} {8}},\ \bibinfo {pages} {031089}
  (\bibinfo {year} {2018})}\BibitemShut {NoStop}%
\bibitem [{\citenamefont {Koshino}\ \emph {et~al.}(2018)\citenamefont
  {Koshino}, \citenamefont {Yuan}, \citenamefont {Koretsune}, \citenamefont
  {Ochi}, \citenamefont {Kuroki},\ and\ \citenamefont {Fu}}]{Koshino2018}%
  \BibitemOpen
  \bibfield  {author} {\bibinfo {author} {\bibfnamefont {M.}~\bibnamefont
  {Koshino}}, \bibinfo {author} {\bibfnamefont {N.~F.~Q.}\ \bibnamefont
  {Yuan}}, \bibinfo {author} {\bibfnamefont {T.}~\bibnamefont {Koretsune}},
  \bibinfo {author} {\bibfnamefont {M.}~\bibnamefont {Ochi}}, \bibinfo {author}
  {\bibfnamefont {K.}~\bibnamefont {Kuroki}}, \ and\ \bibinfo {author}
  {\bibfnamefont {L.}~\bibnamefont {Fu}},\ }\href {\doibase
  10.1103/PhysRevX.8.031087} {\bibfield  {journal} {\bibinfo  {journal} {Phys.
  Rev. X}\ }\textbf {\bibinfo {volume} {8}},\ \bibinfo {pages} {031087}
  (\bibinfo {year} {2018})}\BibitemShut {NoStop}%
\bibitem [{\citenamefont {Zou}\ \emph {et~al.}(2018)\citenamefont {Zou},
  \citenamefont {Po}, \citenamefont {Vishwanath},\ and\ \citenamefont
  {Senthil}}]{Zou2018}%
  \BibitemOpen
  \bibfield  {author} {\bibinfo {author} {\bibfnamefont {L.}~\bibnamefont
  {Zou}}, \bibinfo {author} {\bibfnamefont {H.~C.}\ \bibnamefont {Po}},
  \bibinfo {author} {\bibfnamefont {A.}~\bibnamefont {Vishwanath}}, \ and\
  \bibinfo {author} {\bibfnamefont {T.}~\bibnamefont {Senthil}},\ }\href
  {\doibase 10.1103/PhysRevB.98.085435} {\bibfield  {journal} {\bibinfo
  {journal} {Phys. Rev. B}\ }\textbf {\bibinfo {volume} {98}},\ \bibinfo
  {pages} {085435} (\bibinfo {year} {2018})}\BibitemShut {NoStop}%
\bibitem [{\citenamefont {Kang}\ and\ \citenamefont {Vafek}(2018)}]{Kang2018}%
  \BibitemOpen
  \bibfield  {author} {\bibinfo {author} {\bibfnamefont {J.}~\bibnamefont
  {Kang}}\ and\ \bibinfo {author} {\bibfnamefont {O.}~\bibnamefont {Vafek}},\
  }\href {\doibase 10.1103/PhysRevX.8.031088} {\bibfield  {journal} {\bibinfo
  {journal} {Physical Review X}\ }\textbf {\bibinfo {volume} {8}},\ \bibinfo
  {pages} {031088} (\bibinfo {year} {2018})}\BibitemShut {NoStop}%
\bibitem [{\citenamefont {Rademaker}\ and\ \citenamefont
  {Mellado}(2018)}]{Rademaker2018}%
  \BibitemOpen
  \bibfield  {author} {\bibinfo {author} {\bibfnamefont {L.}~\bibnamefont
  {Rademaker}}\ and\ \bibinfo {author} {\bibfnamefont {P.}~\bibnamefont
  {Mellado}},\ }\href {\doibase 10.1103/PhysRevB.98.235158} {\bibfield
  {journal} {\bibinfo  {journal} {Phys. Rev. B}\ }\textbf {\bibinfo {volume}
  {98}},\ \bibinfo {pages} {235158} (\bibinfo {year} {2018})}\BibitemShut
  {NoStop}%
\bibitem [{\citenamefont {Zhang}\ \emph
  {et~al.}(2019{\natexlab{a}})\citenamefont {Zhang}, \citenamefont {Mao},
  \citenamefont {Cao}, \citenamefont {Jarillo-Herrero},\ and\ \citenamefont
  {Senthil}}]{Zhang2019a}%
  \BibitemOpen
  \bibfield  {author} {\bibinfo {author} {\bibfnamefont {Y.-H.}\ \bibnamefont
  {Zhang}}, \bibinfo {author} {\bibfnamefont {D.}~\bibnamefont {Mao}}, \bibinfo
  {author} {\bibfnamefont {Y.}~\bibnamefont {Cao}}, \bibinfo {author}
  {\bibfnamefont {P.}~\bibnamefont {Jarillo-Herrero}}, \ and\ \bibinfo {author}
  {\bibfnamefont {T.}~\bibnamefont {Senthil}},\ }\href {\doibase
  10.1103/PhysRevB.99.075127} {\bibfield  {journal} {\bibinfo  {journal} {Phys.
  Rev. B}\ }\textbf {\bibinfo {volume} {99}},\ \bibinfo {pages} {075127}
  (\bibinfo {year} {2019}{\natexlab{a}})}\BibitemShut {NoStop}%
\bibitem [{\citenamefont {Zhang}(2019)}]{Zhang2019b}%
  \BibitemOpen
  \bibfield  {author} {\bibinfo {author} {\bibfnamefont {L.}~\bibnamefont
  {Zhang}},\ }\href {\doibase 10.1016/j.scib.2019.03.010} {\bibfield  {journal}
  {\bibinfo  {journal} {Science Bulletin}\ }\textbf {\bibinfo {volume} {64}},\
  \bibinfo {pages} {495} (\bibinfo {year} {2019})}\BibitemShut {NoStop}%
\bibitem [{\citenamefont {Lian}\ \emph {et~al.}(2019)\citenamefont {Lian},
  \citenamefont {Wang},\ and\ \citenamefont {Bernevig}}]{Lian2019}%
  \BibitemOpen
  \bibfield  {author} {\bibinfo {author} {\bibfnamefont {B.}~\bibnamefont
  {Lian}}, \bibinfo {author} {\bibfnamefont {Z.}~\bibnamefont {Wang}}, \ and\
  \bibinfo {author} {\bibfnamefont {B.~A.}\ \bibnamefont {Bernevig}},\ }\href
  {\doibase 10.1103/PhysRevLett.122.257002} {\bibfield  {journal} {\bibinfo
  {journal} {Phys. Rev. Lett.}\ }\textbf {\bibinfo {volume} {122}},\ \bibinfo
  {pages} {257002} (\bibinfo {year} {2019})}\BibitemShut {NoStop}%
\bibitem [{\citenamefont {Wu}\ and\ \citenamefont
  {Das~Sarma}(2019)}]{DasSarma2019}%
  \BibitemOpen
  \bibfield  {author} {\bibinfo {author} {\bibfnamefont {F.}~\bibnamefont
  {Wu}}\ and\ \bibinfo {author} {\bibfnamefont {S.}~\bibnamefont {Das~Sarma}},\
  }\href {\doibase 10.1103/PhysRevB.99.220507} {\bibfield  {journal} {\bibinfo
  {journal} {Phys. Rev. B}\ }\textbf {\bibinfo {volume} {99}},\ \bibinfo
  {pages} {220507} (\bibinfo {year} {2019})}\BibitemShut {NoStop}%
\bibitem [{\citenamefont {Lin}\ and\ \citenamefont
  {Nandkishore}(2019)}]{Nandkishore2019}%
  \BibitemOpen
  \bibfield  {author} {\bibinfo {author} {\bibfnamefont {Y.-P.}\ \bibnamefont
  {Lin}}\ and\ \bibinfo {author} {\bibfnamefont {R.~M.}\ \bibnamefont
  {Nandkishore}},\ }\href {\doibase 10.1103/PhysRevB.100.085136} {\bibfield
  {journal} {\bibinfo  {journal} {Phys. Rev. B}\ }\textbf {\bibinfo {volume}
  {100}},\ \bibinfo {pages} {085136} (\bibinfo {year} {2019})}\BibitemShut
  {NoStop}%
\bibitem [{\citenamefont {Song}\ \emph {et~al.}(2019)\citenamefont {Song},
  \citenamefont {Wang}, \citenamefont {Shi}, \citenamefont {Li}, \citenamefont
  {Fang},\ and\ \citenamefont {Bernevig}}]{Song2019}%
  \BibitemOpen
  \bibfield  {author} {\bibinfo {author} {\bibfnamefont {Z.}~\bibnamefont
  {Song}}, \bibinfo {author} {\bibfnamefont {Z.}~\bibnamefont {Wang}}, \bibinfo
  {author} {\bibfnamefont {W.}~\bibnamefont {Shi}}, \bibinfo {author}
  {\bibfnamefont {G.}~\bibnamefont {Li}}, \bibinfo {author} {\bibfnamefont
  {C.}~\bibnamefont {Fang}}, \ and\ \bibinfo {author} {\bibfnamefont {B.~A.}\
  \bibnamefont {Bernevig}},\ }\href {\doibase 10.1103/PhysRevLett.123.036401}
  {\bibfield  {journal} {\bibinfo  {journal} {Physical Review Letters}\
  }\textbf {\bibinfo {volume} {123}},\ \bibinfo {pages} {036401} (\bibinfo
  {year} {2019})}\BibitemShut {NoStop}%
\bibitem [{\citenamefont {Kang}\ and\ \citenamefont {Vafek}(2019)}]{Kang2019}%
  \BibitemOpen
  \bibfield  {author} {\bibinfo {author} {\bibfnamefont {J.}~\bibnamefont
  {Kang}}\ and\ \bibinfo {author} {\bibfnamefont {O.}~\bibnamefont {Vafek}},\
  }\href {\doibase 10.1103/PhysRevLett.122.246401} {\bibfield  {journal}
  {\bibinfo  {journal} {Phys. Rev. Lett.}\ }\textbf {\bibinfo {volume} {122}},\
  \bibinfo {pages} {246401} (\bibinfo {year} {2019})}\BibitemShut {NoStop}%
\bibitem [{\citenamefont {Tarnopolsky}\ \emph {et~al.}(2019)\citenamefont
  {Tarnopolsky}, \citenamefont {Kruchkov},\ and\ \citenamefont
  {Vishwanath}}]{Tarnopolsky2019}%
  \BibitemOpen
  \bibfield  {author} {\bibinfo {author} {\bibfnamefont {G.}~\bibnamefont
  {Tarnopolsky}}, \bibinfo {author} {\bibfnamefont {A.~J.}\ \bibnamefont
  {Kruchkov}}, \ and\ \bibinfo {author} {\bibfnamefont {A.}~\bibnamefont
  {Vishwanath}},\ }\href {\doibase 10.1103/PhysRevLett.122.106405} {\bibfield
  {journal} {\bibinfo  {journal} {Phys. Rev. Lett.}\ }\textbf {\bibinfo
  {volume} {122}},\ \bibinfo {pages} {106405} (\bibinfo {year}
  {2019})}\BibitemShut {NoStop}%
\bibitem [{\citenamefont {Xie}\ \emph {et~al.}(2019)\citenamefont {Xie},
  \citenamefont {Lian}, \citenamefont {J{\"a}ck}, \citenamefont {Liu},
  \citenamefont {Chiu}, \citenamefont {Watanabe}, \citenamefont {Taniguchi},
  \citenamefont {Bernevig},\ and\ \citenamefont {Yazdani}}]{STM_Yazdani19}%
  \BibitemOpen
  \bibfield  {author} {\bibinfo {author} {\bibfnamefont {Y.}~\bibnamefont
  {Xie}}, \bibinfo {author} {\bibfnamefont {B.}~\bibnamefont {Lian}}, \bibinfo
  {author} {\bibfnamefont {B.}~\bibnamefont {J{\"a}ck}}, \bibinfo {author}
  {\bibfnamefont {X.}~\bibnamefont {Liu}}, \bibinfo {author} {\bibfnamefont
  {C.-L.}\ \bibnamefont {Chiu}}, \bibinfo {author} {\bibfnamefont
  {K.}~\bibnamefont {Watanabe}}, \bibinfo {author} {\bibfnamefont
  {T.}~\bibnamefont {Taniguchi}}, \bibinfo {author} {\bibfnamefont {B.~A.}\
  \bibnamefont {Bernevig}}, \ and\ \bibinfo {author} {\bibfnamefont
  {A.}~\bibnamefont {Yazdani}},\ }\href@noop {} {\bibfield  {journal} {\bibinfo
   {journal} {Nature}\ }\textbf {\bibinfo {volume} {572}},\ \bibinfo {pages}
  {101} (\bibinfo {year} {2019})}\BibitemShut {NoStop}%
\bibitem [{\citenamefont {Choi}\ \emph {et~al.}(2019)\citenamefont {Choi},
  \citenamefont {Kemmer}, \citenamefont {Peng}, \citenamefont {Thomson},
  \citenamefont {Arora}, \citenamefont {Polski}, \citenamefont {Zhang},
  \citenamefont {Ren}, \citenamefont {Alicea}, \citenamefont {Refael},
  \citenamefont {Oppen}, \citenamefont {Watanabe}, \citenamefont {Taniguchi},\
  and\ \citenamefont {Nadj-Perge}}]{STM_Perge19}%
  \BibitemOpen
  \bibfield  {author} {\bibinfo {author} {\bibfnamefont {Y.}~\bibnamefont
  {Choi}}, \bibinfo {author} {\bibfnamefont {J.}~\bibnamefont {Kemmer}},
  \bibinfo {author} {\bibfnamefont {Y.}~\bibnamefont {Peng}}, \bibinfo {author}
  {\bibfnamefont {A.}~\bibnamefont {Thomson}}, \bibinfo {author} {\bibfnamefont
  {H.}~\bibnamefont {Arora}}, \bibinfo {author} {\bibfnamefont
  {R.}~\bibnamefont {Polski}}, \bibinfo {author} {\bibfnamefont
  {Y.}~\bibnamefont {Zhang}}, \bibinfo {author} {\bibfnamefont
  {H.}~\bibnamefont {Ren}}, \bibinfo {author} {\bibfnamefont {J.}~\bibnamefont
  {Alicea}}, \bibinfo {author} {\bibfnamefont {G.}~\bibnamefont {Refael}},
  \bibinfo {author} {\bibfnamefont {F.~v.}\ \bibnamefont {Oppen}}, \bibinfo
  {author} {\bibfnamefont {K.}~\bibnamefont {Watanabe}}, \bibinfo {author}
  {\bibfnamefont {T.}~\bibnamefont {Taniguchi}}, \ and\ \bibinfo {author}
  {\bibfnamefont {S.}~\bibnamefont {Nadj-Perge}},\ }\href@noop {} {\bibfield
  {journal} {\bibinfo  {journal} {Nature Physics}\ }\textbf {\bibinfo {volume}
  {15}},\ \bibinfo {pages} {1174} (\bibinfo {year} {2019})}\BibitemShut
  {NoStop}%
\bibitem [{\citenamefont {Kerelsky}\ \emph {et~al.}(2019)\citenamefont
  {Kerelsky}, \citenamefont {McGilly}, \citenamefont {Kennes}, \citenamefont
  {Xian}, \citenamefont {Yankowitz}, \citenamefont {Chen}, \citenamefont
  {Watanabe}, \citenamefont {Taniguchi}, \citenamefont {Hone}, \citenamefont
  {Dean}, \citenamefont {Rubio},\ and\ \citenamefont
  {Pasupathy}}]{STM_Pasupathy19}%
  \BibitemOpen
  \bibfield  {author} {\bibinfo {author} {\bibfnamefont {A.}~\bibnamefont
  {Kerelsky}}, \bibinfo {author} {\bibfnamefont {L.~J.}\ \bibnamefont
  {McGilly}}, \bibinfo {author} {\bibfnamefont {D.~M.}\ \bibnamefont {Kennes}},
  \bibinfo {author} {\bibfnamefont {L.}~\bibnamefont {Xian}}, \bibinfo {author}
  {\bibfnamefont {M.}~\bibnamefont {Yankowitz}}, \bibinfo {author}
  {\bibfnamefont {S.}~\bibnamefont {Chen}}, \bibinfo {author} {\bibfnamefont
  {K.}~\bibnamefont {Watanabe}}, \bibinfo {author} {\bibfnamefont
  {T.}~\bibnamefont {Taniguchi}}, \bibinfo {author} {\bibfnamefont
  {J.}~\bibnamefont {Hone}}, \bibinfo {author} {\bibfnamefont {C.}~\bibnamefont
  {Dean}}, \bibinfo {author} {\bibfnamefont {A.}~\bibnamefont {Rubio}}, \ and\
  \bibinfo {author} {\bibfnamefont {A.~N.}\ \bibnamefont {Pasupathy}},\
  }\href@noop {} {\bibfield  {journal} {\bibinfo  {journal} {Nature}\ }\textbf
  {\bibinfo {volume} {572}},\ \bibinfo {pages} {95} (\bibinfo {year}
  {2019})}\BibitemShut {NoStop}%
\bibitem [{\citenamefont {Jiang}\ \emph {et~al.}(2019)\citenamefont {Jiang},
  \citenamefont {Lai}, \citenamefont {Watanabe}, \citenamefont {Taniguchi},
  \citenamefont {Haule}, \citenamefont {Mao},\ and\ \citenamefont
  {Andrei}}]{STM_Andrei19}%
  \BibitemOpen
  \bibfield  {author} {\bibinfo {author} {\bibfnamefont {Y.}~\bibnamefont
  {Jiang}}, \bibinfo {author} {\bibfnamefont {X.}~\bibnamefont {Lai}}, \bibinfo
  {author} {\bibfnamefont {K.}~\bibnamefont {Watanabe}}, \bibinfo {author}
  {\bibfnamefont {T.}~\bibnamefont {Taniguchi}}, \bibinfo {author}
  {\bibfnamefont {K.}~\bibnamefont {Haule}}, \bibinfo {author} {\bibfnamefont
  {J.}~\bibnamefont {Mao}}, \ and\ \bibinfo {author} {\bibfnamefont {E.~Y.}\
  \bibnamefont {Andrei}},\ }\href@noop {} {\bibfield  {journal} {\bibinfo
  {journal} {Nature}\ }\textbf {\bibinfo {volume} {573}},\ \bibinfo {pages}
  {91} (\bibinfo {year} {2019})}\BibitemShut {NoStop}%
\bibitem [{\citenamefont {Jarillo-Herrero}(2019)}]{Pablo_nematics}%
  \BibitemOpen
  \bibfield  {author} {\bibinfo {author} {\bibfnamefont {P.}~\bibnamefont
  {Jarillo-Herrero}},\ }\href@noop {} {\bibfield  {journal} {\bibinfo
  {journal} {KITP Workshop: Correlations in Moir{\'e} Flat Bands}\ } (\bibinfo
  {year} {2019})}\BibitemShut {NoStop}%
\bibitem [{\citenamefont {Zhang}\ \emph
  {et~al.}(2019{\natexlab{b}})\citenamefont {Zhang}, \citenamefont {Po},\ and\
  \citenamefont {Senthil}}]{Senthil19_C3z}%
  \BibitemOpen
  \bibfield  {author} {\bibinfo {author} {\bibfnamefont {Y.-H.}\ \bibnamefont
  {Zhang}}, \bibinfo {author} {\bibfnamefont {H.~C.}\ \bibnamefont {Po}}, \
  and\ \bibinfo {author} {\bibfnamefont {T.}~\bibnamefont {Senthil}},\
  }\href@noop {} {\bibfield  {journal} {\bibinfo  {journal} {arXiv:1904.10452}\
  } (\bibinfo {year} {2019}{\natexlab{b}})}\BibitemShut {NoStop}%
\bibitem [{\citenamefont {Liu}\ \emph {et~al.}(2019{\natexlab{b}})\citenamefont
  {Liu}, \citenamefont {Khalaf}, \citenamefont {Lee},\ and\ \citenamefont
  {Vishwanath}}]{Vishwanath19_nematic}%
  \BibitemOpen
  \bibfield  {author} {\bibinfo {author} {\bibfnamefont {S.}~\bibnamefont
  {Liu}}, \bibinfo {author} {\bibfnamefont {E.}~\bibnamefont {Khalaf}},
  \bibinfo {author} {\bibfnamefont {J.~Y.}\ \bibnamefont {Lee}}, \ and\
  \bibinfo {author} {\bibfnamefont {A.}~\bibnamefont {Vishwanath}},\
  }\href@noop {} {\bibfield  {journal} {\bibinfo  {journal} {arXiv:1905.07409}\
  } (\bibinfo {year} {2019}{\natexlab{b}})}\BibitemShut {NoStop}%
\bibitem [{\citenamefont {Fradkin}\ \emph {et~al.}(2010)\citenamefont
  {Fradkin}, \citenamefont {Kivelson}, \citenamefont {Lawler}, \citenamefont
  {Eisenstein},\ and\ \citenamefont {Mackenzie}}]{Fradkin_review}%
  \BibitemOpen
  \bibfield  {author} {\bibinfo {author} {\bibfnamefont {E.}~\bibnamefont
  {Fradkin}}, \bibinfo {author} {\bibfnamefont {S.~A.}\ \bibnamefont
  {Kivelson}}, \bibinfo {author} {\bibfnamefont {M.~J.}\ \bibnamefont
  {Lawler}}, \bibinfo {author} {\bibfnamefont {J.~P.}\ \bibnamefont
  {Eisenstein}}, \ and\ \bibinfo {author} {\bibfnamefont {A.~P.}\ \bibnamefont
  {Mackenzie}},\ }\href {\doibase 10.1146/annurev-conmatphys-070909-103925}
  {\bibfield  {journal} {\bibinfo  {journal} {Annual Review of Condensed Matter
  Physics}\ }\textbf {\bibinfo {volume} {1}},\ \bibinfo {pages} {153} (\bibinfo
  {year} {2010})}\BibitemShut {NoStop}%
\bibitem [{\citenamefont {Fernandes}\ \emph {et~al.}(2019)\citenamefont
  {Fernandes}, \citenamefont {Orth},\ and\ \citenamefont
  {Schmalian}}]{Fernandes_review19}%
  \BibitemOpen
  \bibfield  {author} {\bibinfo {author} {\bibfnamefont {R.~M.}\ \bibnamefont
  {Fernandes}}, \bibinfo {author} {\bibfnamefont {P.~P.}\ \bibnamefont {Orth}},
  \ and\ \bibinfo {author} {\bibfnamefont {J.}~\bibnamefont {Schmalian}},\
  }\href {\doibase 10.1146/annurev-conmatphys-031218-013200} {\bibfield
  {journal} {\bibinfo  {journal} {Annual Review of Condensed Matter Physics}\
  }\textbf {\bibinfo {volume} {10}},\ \bibinfo {pages} {133} (\bibinfo {year}
  {2019})}\BibitemShut {NoStop}%
\bibitem [{\citenamefont {Venderbos}\ and\ \citenamefont
  {Fernandes}(2018)}]{Venderbos18}%
  \BibitemOpen
  \bibfield  {author} {\bibinfo {author} {\bibfnamefont {J.~W.~F.}\
  \bibnamefont {Venderbos}}\ and\ \bibinfo {author} {\bibfnamefont {R.~M.}\
  \bibnamefont {Fernandes}},\ }\href {\doibase 10.1103/PhysRevB.98.245103}
  {\bibfield  {journal} {\bibinfo  {journal} {Phys. Rev. B}\ }\textbf {\bibinfo
  {volume} {98}},\ \bibinfo {pages} {245103} (\bibinfo {year}
  {2018})}\BibitemShut {NoStop}%
\bibitem [{\citenamefont {Dodaro}\ \emph {et~al.}(2018)\citenamefont {Dodaro},
  \citenamefont {Kivelson}, \citenamefont {Schattner}, \citenamefont {Sun},\
  and\ \citenamefont {Wang}}]{Dodaro2018}%
  \BibitemOpen
  \bibfield  {author} {\bibinfo {author} {\bibfnamefont {J.~F.}\ \bibnamefont
  {Dodaro}}, \bibinfo {author} {\bibfnamefont {S.~A.}\ \bibnamefont
  {Kivelson}}, \bibinfo {author} {\bibfnamefont {Y.}~\bibnamefont {Schattner}},
  \bibinfo {author} {\bibfnamefont {X.~Q.}\ \bibnamefont {Sun}}, \ and\
  \bibinfo {author} {\bibfnamefont {C.}~\bibnamefont {Wang}},\ }\href {\doibase
  10.1103/PhysRevB.98.075154} {\bibfield  {journal} {\bibinfo  {journal} {Phys.
  Rev. B}\ }\textbf {\bibinfo {volume} {98}},\ \bibinfo {pages} {075154}
  (\bibinfo {year} {2018})}\BibitemShut {NoStop}%
\bibitem [{\citenamefont {Isobe}\ \emph {et~al.}(2018)\citenamefont {Isobe},
  \citenamefont {Yuan},\ and\ \citenamefont {Fu}}]{Isobe2018}%
  \BibitemOpen
  \bibfield  {author} {\bibinfo {author} {\bibfnamefont {H.}~\bibnamefont
  {Isobe}}, \bibinfo {author} {\bibfnamefont {N.~F.~Q.}\ \bibnamefont {Yuan}},
  \ and\ \bibinfo {author} {\bibfnamefont {L.}~\bibnamefont {Fu}},\ }\href
  {\doibase 10.1103/PhysRevX.8.041041} {\bibfield  {journal} {\bibinfo
  {journal} {Phys. Rev. X}\ }\textbf {\bibinfo {volume} {8}},\ \bibinfo {pages}
  {041041} (\bibinfo {year} {2018})}\BibitemShut {NoStop}%
\bibitem [{\citenamefont {Kozii}\ \emph {et~al.}(2019)\citenamefont {Kozii},
  \citenamefont {Isobe}, \citenamefont {Venderbos},\ and\ \citenamefont
  {Fu}}]{Kozii2019}%
  \BibitemOpen
  \bibfield  {author} {\bibinfo {author} {\bibfnamefont {V.}~\bibnamefont
  {Kozii}}, \bibinfo {author} {\bibfnamefont {H.}~\bibnamefont {Isobe}},
  \bibinfo {author} {\bibfnamefont {J.~W.~F.}\ \bibnamefont {Venderbos}}, \
  and\ \bibinfo {author} {\bibfnamefont {L.}~\bibnamefont {Fu}},\ }\href
  {\doibase 10.1103/PhysRevB.99.144507} {\bibfield  {journal} {\bibinfo
  {journal} {Phys. Rev. B}\ }\textbf {\bibinfo {volume} {99}},\ \bibinfo
  {pages} {144507} (\bibinfo {year} {2019})}\BibitemShut {NoStop}%
\bibitem [{\citenamefont {Chichinadze}\ \emph {et~al.}(2019)\citenamefont
  {Chichinadze}, \citenamefont {Classen},\ and\ \citenamefont
  {Chubukov}}]{Chubukov2019}%
  \BibitemOpen
  \bibfield  {author} {\bibinfo {author} {\bibfnamefont {D.~V.}\ \bibnamefont
  {Chichinadze}}, \bibinfo {author} {\bibfnamefont {L.}~\bibnamefont
  {Classen}}, \ and\ \bibinfo {author} {\bibfnamefont {A.~V.}\ \bibnamefont
  {Chubukov}},\ }\href@noop {} {\bibfield  {journal} {\bibinfo  {journal}
  {arXiv:1910.07379}\ } (\bibinfo {year} {2019})}\BibitemShut {NoStop}%
\bibitem [{\citenamefont {Uri}\ \emph {et~al.}(2019)\citenamefont {Uri},
  \citenamefont {Grover}, \citenamefont {Cao}, \citenamefont {Crosse},
  \citenamefont {Bagani}, \citenamefont {Rodan-Legrain}, \citenamefont
  {Myasoedov}, \citenamefont {Watanabe}, \citenamefont {Taniguchi},
  \citenamefont {Moon}, \citenamefont {Koshino}, \citenamefont
  {Jarillo-Herrero},\ and\ \citenamefont {Zeldov}}]{Zeldov19}%
  \BibitemOpen
  \bibfield  {author} {\bibinfo {author} {\bibfnamefont {A.}~\bibnamefont
  {Uri}}, \bibinfo {author} {\bibfnamefont {S.}~\bibnamefont {Grover}},
  \bibinfo {author} {\bibfnamefont {Y.}~\bibnamefont {Cao}}, \bibinfo {author}
  {\bibfnamefont {J.}~\bibnamefont {Crosse}}, \bibinfo {author} {\bibfnamefont
  {K.}~\bibnamefont {Bagani}}, \bibinfo {author} {\bibfnamefont
  {D.}~\bibnamefont {Rodan-Legrain}}, \bibinfo {author} {\bibfnamefont
  {Y.}~\bibnamefont {Myasoedov}}, \bibinfo {author} {\bibfnamefont
  {K.}~\bibnamefont {Watanabe}}, \bibinfo {author} {\bibfnamefont
  {T.}~\bibnamefont {Taniguchi}}, \bibinfo {author} {\bibfnamefont
  {P.}~\bibnamefont {Moon}}, \bibinfo {author} {\bibfnamefont {M.}~\bibnamefont
  {Koshino}}, \bibinfo {author} {\bibfnamefont {P.}~\bibnamefont
  {Jarillo-Herrero}}, \ and\ \bibinfo {author} {\bibfnamefont {E.}~\bibnamefont
  {Zeldov}},\ }\href@noop {} {\bibfield  {journal} {\bibinfo  {journal}
  {arXiv:1908.04595}\ } (\bibinfo {year} {2019})}\BibitemShut {NoStop}%
\bibitem [{\citenamefont {Cea}\ \emph {et~al.}(2019)\citenamefont {Cea},
  \citenamefont {Waler},\ and\ \citenamefont {Guinea}}]{Guinea19_strain}%
  \BibitemOpen
  \bibfield  {author} {\bibinfo {author} {\bibfnamefont {T.}~\bibnamefont
  {Cea}}, \bibinfo {author} {\bibfnamefont {N.~R.}\ \bibnamefont {Waler}}, \
  and\ \bibinfo {author} {\bibfnamefont {F.}~\bibnamefont {Guinea}},\
  }\href@noop {} {\bibfield  {journal} {\bibinfo  {journal} {arXiv:1906.10570}\
  } (\bibinfo {year} {2019})}\BibitemShut {NoStop}%
\bibitem [{\citenamefont {Wilson}\ \emph {et~al.}(2019)\citenamefont {Wilson},
  \citenamefont {Fu}, \citenamefont {Das~Sarma},\ and\ \citenamefont
  {Pixley}}]{Pixley2019}%
  \BibitemOpen
  \bibfield  {author} {\bibinfo {author} {\bibfnamefont {J.~H.}\ \bibnamefont
  {Wilson}}, \bibinfo {author} {\bibfnamefont {Y.}~\bibnamefont {Fu}}, \bibinfo
  {author} {\bibfnamefont {S.}~\bibnamefont {Das~Sarma}}, \ and\ \bibinfo
  {author} {\bibfnamefont {J.~H.}\ \bibnamefont {Pixley}},\ }\href@noop {}
  {\bibfield  {journal} {\bibinfo  {journal} {arXiv:1908.02753}\ } (\bibinfo
  {year} {2019})}\BibitemShut {NoStop}%
\bibitem [{\citenamefont {Fernandes}\ \emph {et~al.}(2014)\citenamefont
  {Fernandes}, \citenamefont {Chubukov},\ and\ \citenamefont
  {Schmalian}}]{Fernandes14}%
  \BibitemOpen
  \bibfield  {author} {\bibinfo {author} {\bibfnamefont {R.~M.}\ \bibnamefont
  {Fernandes}}, \bibinfo {author} {\bibfnamefont {A.~V.}\ \bibnamefont
  {Chubukov}}, \ and\ \bibinfo {author} {\bibfnamefont {J.}~\bibnamefont
  {Schmalian}},\ }\href@noop {} {\bibfield  {journal} {\bibinfo  {journal}
  {Nature Physics}\ }\textbf {\bibinfo {volume} {10}},\ \bibinfo {pages} {97}
  (\bibinfo {year} {2014})}\BibitemShut {NoStop}%
\bibitem [{\citenamefont {Koshino}\ and\ \citenamefont
  {Son}(2019)}]{Koshino_phonons19}%
  \BibitemOpen
  \bibfield  {author} {\bibinfo {author} {\bibfnamefont {M.}~\bibnamefont
  {Koshino}}\ and\ \bibinfo {author} {\bibfnamefont {Y.-W.}\ \bibnamefont
  {Son}},\ }\href {\doibase 10.1103/PhysRevB.100.075416} {\bibfield  {journal}
  {\bibinfo  {journal} {Phys. Rev. B}\ }\textbf {\bibinfo {volume} {100}},\
  \bibinfo {pages} {075416} (\bibinfo {year} {2019})}\BibitemShut {NoStop}%
\bibitem [{\citenamefont {Ochoa}(2019)}]{Ochoa19}%
  \BibitemOpen
  \bibfield  {author} {\bibinfo {author} {\bibfnamefont {H.}~\bibnamefont
  {Ochoa}},\ }\href {\doibase 10.1103/PhysRevB.100.155426} {\bibfield
  {journal} {\bibinfo  {journal} {Phys. Rev. B}\ }\textbf {\bibinfo {volume}
  {100}},\ \bibinfo {pages} {155426} (\bibinfo {year} {2019})}\BibitemShut
  {NoStop}%
\bibitem [{Note1()}]{Note1}%
  \BibitemOpen
  \bibinfo {note} {The main results presented here also hold if one instead
  considers the approach in which TBG is described in terms of the $D_{3}$
  point group.}\BibitemShut {Stop}%
\bibitem [{\citenamefont {Hecker}\ and\ \citenamefont
  {Schmalian}(2018)}]{Hecker18}%
  \BibitemOpen
  \bibfield  {author} {\bibinfo {author} {\bibfnamefont {M.}~\bibnamefont
  {Hecker}}\ and\ \bibinfo {author} {\bibfnamefont {J.}~\bibnamefont
  {Schmalian}},\ }\href@noop {} {\bibfield  {journal} {\bibinfo  {journal} {npj
  Quantum Materials}\ }\textbf {\bibinfo {volume} {3}},\ \bibinfo {pages} {26}
  (\bibinfo {year} {2018})}\BibitemShut {NoStop}%
\bibitem [{\citenamefont {Little}\ \emph {et~al.}(2019)\citenamefont {Little},
  \citenamefont {Lee}, \citenamefont {John}, \citenamefont {Doyle},
  \citenamefont {Maniv}, \citenamefont {Nair}, \citenamefont {Chen},
  \citenamefont {Rees}, \citenamefont {Venderbos}, \citenamefont {Fernandes},
  \citenamefont {Analytis},\ and\ \citenamefont {Orenstein}}]{Orenstein19}%
  \BibitemOpen
  \bibfield  {author} {\bibinfo {author} {\bibfnamefont {A.}~\bibnamefont
  {Little}}, \bibinfo {author} {\bibfnamefont {C.}~\bibnamefont {Lee}},
  \bibinfo {author} {\bibfnamefont {C.}~\bibnamefont {John}}, \bibinfo {author}
  {\bibfnamefont {S.}~\bibnamefont {Doyle}}, \bibinfo {author} {\bibfnamefont
  {E.}~\bibnamefont {Maniv}}, \bibinfo {author} {\bibfnamefont {N.~L.}\
  \bibnamefont {Nair}}, \bibinfo {author} {\bibfnamefont {W.}~\bibnamefont
  {Chen}}, \bibinfo {author} {\bibfnamefont {D.}~\bibnamefont {Rees}}, \bibinfo
  {author} {\bibfnamefont {J.~W.}\ \bibnamefont {Venderbos}}, \bibinfo {author}
  {\bibfnamefont {R.}~\bibnamefont {Fernandes}}, \bibinfo {author}
  {\bibfnamefont {J.~G.}\ \bibnamefont {Analytis}}, \ and\ \bibinfo {author}
  {\bibfnamefont {J.}~\bibnamefont {Orenstein}},\ }\href@noop {} {\bibfield
  {journal} {\bibinfo  {journal} {arXiv:1908.00657}\ } (\bibinfo {year}
  {2019})}\BibitemShut {NoStop}%
\bibitem [{\citenamefont {Jin}\ \emph {et~al.}(2019)\citenamefont {Jin},
  \citenamefont {Zhang}, \citenamefont {Guo}, \citenamefont {Chen},
  \citenamefont {Zhou},\ and\ \citenamefont {Li}}]{Li_cold_atoms}%
  \BibitemOpen
  \bibfield  {author} {\bibinfo {author} {\bibfnamefont {S.}~\bibnamefont
  {Jin}}, \bibinfo {author} {\bibfnamefont {W.}~\bibnamefont {Zhang}}, \bibinfo
  {author} {\bibfnamefont {X.}~\bibnamefont {Guo}}, \bibinfo {author}
  {\bibfnamefont {X.}~\bibnamefont {Chen}}, \bibinfo {author} {\bibfnamefont
  {X.}~\bibnamefont {Zhou}}, \ and\ \bibinfo {author} {\bibfnamefont
  {X.}~\bibnamefont {Li}},\ }\href@noop {} {\bibfield  {journal} {\bibinfo
  {journal} {arXiv:1910.11880}\ } (\bibinfo {year} {2019})}\BibitemShut
  {NoStop}%
\bibitem [{\citenamefont {Wu}(1982)}]{Yu_Potts82}%
  \BibitemOpen
  \bibfield  {author} {\bibinfo {author} {\bibfnamefont {F.~Y.}\ \bibnamefont
  {Wu}},\ }\href {\doibase 10.1103/RevModPhys.54.235} {\bibfield  {journal}
  {\bibinfo  {journal} {Rev. Mod. Phys.}\ }\textbf {\bibinfo {volume} {54}},\
  \bibinfo {pages} {235} (\bibinfo {year} {1982})}\BibitemShut {NoStop}%
\bibitem [{\citenamefont {Blankschtein}\ and\ \citenamefont
  {Aharony}(1980)}]{Aharony80}%
  \BibitemOpen
  \bibfield  {author} {\bibinfo {author} {\bibfnamefont {D.}~\bibnamefont
  {Blankschtein}}\ and\ \bibinfo {author} {\bibfnamefont {A.}~\bibnamefont
  {Aharony}},\ }\href {\doibase 10.1088/0022-3719/13/25/007} {\bibfield
  {journal} {\bibinfo  {journal} {Journal of Physics C: Solid State Physics}\
  }\textbf {\bibinfo {volume} {13}},\ \bibinfo {pages} {4635} (\bibinfo {year}
  {1980})}\BibitemShut {NoStop}%
\bibitem [{\citenamefont {Karahasanovic}\ and\ \citenamefont
  {Schmalian}(2016)}]{Schmalian16}%
  \BibitemOpen
  \bibfield  {author} {\bibinfo {author} {\bibfnamefont {U.}~\bibnamefont
  {Karahasanovic}}\ and\ \bibinfo {author} {\bibfnamefont {J.}~\bibnamefont
  {Schmalian}},\ }\href {\doibase 10.1103/PhysRevB.93.064520} {\bibfield
  {journal} {\bibinfo  {journal} {Phys. Rev. B}\ }\textbf {\bibinfo {volume}
  {93}},\ \bibinfo {pages} {064520} (\bibinfo {year} {2016})}\BibitemShut
  {NoStop}%
\bibitem [{\citenamefont {Paul}\ and\ \citenamefont {Garst}(2017)}]{Paul17}%
  \BibitemOpen
  \bibfield  {author} {\bibinfo {author} {\bibfnamefont {I.}~\bibnamefont
  {Paul}}\ and\ \bibinfo {author} {\bibfnamefont {M.}~\bibnamefont {Garst}},\
  }\href {\doibase 10.1103/PhysRevLett.118.227601} {\bibfield  {journal}
  {\bibinfo  {journal} {Phys. Rev. Lett.}\ }\textbf {\bibinfo {volume} {118}},\
  \bibinfo {pages} {227601} (\bibinfo {year} {2017})}\BibitemShut {NoStop}%
\bibitem [{\citenamefont {de~Carvalho}\ and\ \citenamefont
  {Fernandes}(2019)}]{Carvalho19}%
  \BibitemOpen
  \bibfield  {author} {\bibinfo {author} {\bibfnamefont {V.}~\bibnamefont
  {de~Carvalho}}\ and\ \bibinfo {author} {\bibfnamefont {R.}~\bibnamefont
  {Fernandes}},\ }\href@noop {} {\bibfield  {journal} {\bibinfo  {journal}
  {arXiv:1906.03205}\ } (\bibinfo {year} {2019})}\BibitemShut {NoStop}%
\bibitem [{\citenamefont {Wu}\ \emph {et~al.}(2018)\citenamefont {Wu},
  \citenamefont {MacDonald},\ and\ \citenamefont {Martin}}]{Wu2018}%
  \BibitemOpen
  \bibfield  {author} {\bibinfo {author} {\bibfnamefont {F.}~\bibnamefont
  {Wu}}, \bibinfo {author} {\bibfnamefont {A.~H.}\ \bibnamefont {MacDonald}}, \
  and\ \bibinfo {author} {\bibfnamefont {I.}~\bibnamefont {Martin}},\ }\href
  {\doibase 10.1103/PhysRevLett.121.257001} {\bibfield  {journal} {\bibinfo
  {journal} {Phys. Rev. Lett.}\ }\textbf {\bibinfo {volume} {121}},\ \bibinfo
  {pages} {257001} (\bibinfo {year} {2018})}\BibitemShut {NoStop}%
\bibitem [{\citenamefont {Angeli}\ \emph {et~al.}(2019)\citenamefont {Angeli},
  \citenamefont {Tosatti},\ and\ \citenamefont {Fabrizio}}]{Frabrizio19}%
  \BibitemOpen
  \bibfield  {author} {\bibinfo {author} {\bibfnamefont {M.}~\bibnamefont
  {Angeli}}, \bibinfo {author} {\bibfnamefont {E.}~\bibnamefont {Tosatti}}, \
  and\ \bibinfo {author} {\bibfnamefont {M.}~\bibnamefont {Fabrizio}},\
  }\href@noop {} {\bibfield  {journal} {\bibinfo  {journal} {arXiv preprint
  arXiv:1904.06301}\ } (\bibinfo {year} {2019})}\BibitemShut {NoStop}%
\bibitem [{\citenamefont {Wu}\ \emph {et~al.}(2019)\citenamefont {Wu},
  \citenamefont {Hwang},\ and\ \citenamefont {Das~Sarma}}]{DasSarma_phonons}%
  \BibitemOpen
  \bibfield  {author} {\bibinfo {author} {\bibfnamefont {F.}~\bibnamefont
  {Wu}}, \bibinfo {author} {\bibfnamefont {E.}~\bibnamefont {Hwang}}, \ and\
  \bibinfo {author} {\bibfnamefont {S.}~\bibnamefont {Das~Sarma}},\ }\href
  {\doibase 10.1103/PhysRevB.99.165112} {\bibfield  {journal} {\bibinfo
  {journal} {Phys. Rev. B}\ }\textbf {\bibinfo {volume} {99}},\ \bibinfo
  {pages} {165112} (\bibinfo {year} {2019})}\BibitemShut {NoStop}%
\bibitem [{\citenamefont {Cowley}(1976)}]{Cowley76}%
  \BibitemOpen
  \bibfield  {author} {\bibinfo {author} {\bibfnamefont {R.~A.}\ \bibnamefont
  {Cowley}},\ }\href {\doibase 10.1103/PhysRevB.13.4877} {\bibfield  {journal}
  {\bibinfo  {journal} {Phys. Rev. B}\ }\textbf {\bibinfo {volume} {13}},\
  \bibinfo {pages} {4877} (\bibinfo {year} {1976})}\BibitemShut {NoStop}%
\bibitem [{\citenamefont {Folk}\ \emph {et~al.}(1976)\citenamefont {Folk},
  \citenamefont {Iro},\ and\ \citenamefont {Schwabl}}]{Folk76}%
  \BibitemOpen
  \bibfield  {author} {\bibinfo {author} {\bibfnamefont {R.}~\bibnamefont
  {Folk}}, \bibinfo {author} {\bibfnamefont {H.}~\bibnamefont {Iro}}, \ and\
  \bibinfo {author} {\bibfnamefont {F.}~\bibnamefont {Schwabl}},\ }\href@noop
  {} {\bibfield  {journal} {\bibinfo  {journal} {Zeitschrift f{\"u}r Physik B
  Condensed Matter}\ }\textbf {\bibinfo {volume} {25}},\ \bibinfo {pages} {69}
  (\bibinfo {year} {1976})}\BibitemShut {NoStop}%
\bibitem [{\citenamefont {Po}\ \emph {et~al.}(2019)\citenamefont {Po},
  \citenamefont {Zou}, \citenamefont {Senthil},\ and\ \citenamefont
  {Vishwanath}}]{Po2019}%
  \BibitemOpen
  \bibfield  {author} {\bibinfo {author} {\bibfnamefont {H.~C.}\ \bibnamefont
  {Po}}, \bibinfo {author} {\bibfnamefont {L.}~\bibnamefont {Zou}}, \bibinfo
  {author} {\bibfnamefont {T.}~\bibnamefont {Senthil}}, \ and\ \bibinfo
  {author} {\bibfnamefont {A.}~\bibnamefont {Vishwanath}},\ }\href {\doibase
  10.1103/PhysRevB.99.195455} {\bibfield  {journal} {\bibinfo  {journal} {Phys.
  Rev. B}\ }\textbf {\bibinfo {volume} {99}},\ \bibinfo {pages} {195455}
  (\bibinfo {year} {2019})}\BibitemShut {NoStop}%
\bibitem [{\citenamefont {Valenzuela}\ and\ \citenamefont
  {Vozmediano}(2008)}]{Valenzuela08}%
  \BibitemOpen
  \bibfield  {author} {\bibinfo {author} {\bibfnamefont {B.}~\bibnamefont
  {Valenzuela}}\ and\ \bibinfo {author} {\bibfnamefont {M.~A.~H.}\ \bibnamefont
  {Vozmediano}},\ }\href {\doibase 10.1088/1367-2630/10/11/113009} {\bibfield
  {journal} {\bibinfo  {journal} {New Journal of Physics}\ }\textbf {\bibinfo
  {volume} {10}},\ \bibinfo {pages} {113009} (\bibinfo {year}
  {2008})}\BibitemShut {NoStop}%
\bibitem [{\citenamefont {Xu}\ and\ \citenamefont {Balents}(2018)}]{Xu2018a}%
  \BibitemOpen
  \bibfield  {author} {\bibinfo {author} {\bibfnamefont {C.}~\bibnamefont
  {Xu}}\ and\ \bibinfo {author} {\bibfnamefont {L.}~\bibnamefont {Balents}},\
  }\href {\doibase 10.1103/PhysRevLett.121.087001} {\bibfield  {journal}
  {\bibinfo  {journal} {Phys. Rev. Lett.}\ }\textbf {\bibinfo {volume} {121}},\
  \bibinfo {pages} {087001} (\bibinfo {year} {2018})}\BibitemShut {NoStop}%
\bibitem [{\citenamefont {Classen}\ \emph {et~al.}(2019)\citenamefont
  {Classen}, \citenamefont {Honerkamp},\ and\ \citenamefont
  {Scherer}}]{Classen19}%
  \BibitemOpen
  \bibfield  {author} {\bibinfo {author} {\bibfnamefont {L.}~\bibnamefont
  {Classen}}, \bibinfo {author} {\bibfnamefont {C.}~\bibnamefont {Honerkamp}},
  \ and\ \bibinfo {author} {\bibfnamefont {M.~M.}\ \bibnamefont {Scherer}},\
  }\href {\doibase 10.1103/PhysRevB.99.195120} {\bibfield  {journal} {\bibinfo
  {journal} {Phys. Rev. B}\ }\textbf {\bibinfo {volume} {99}},\ \bibinfo
  {pages} {195120} (\bibinfo {year} {2019})}\BibitemShut {NoStop}%
\bibitem [{\citenamefont {Kiese}\ \emph {et~al.}(2019)\citenamefont {Kiese},
  \citenamefont {Buessen}, \citenamefont {Hickey}, \citenamefont {Trebst},\
  and\ \citenamefont {Scherer}}]{Kiese19}%
  \BibitemOpen
  \bibfield  {author} {\bibinfo {author} {\bibfnamefont {D.}~\bibnamefont
  {Kiese}}, \bibinfo {author} {\bibfnamefont {F.~L.}\ \bibnamefont {Buessen}},
  \bibinfo {author} {\bibfnamefont {C.}~\bibnamefont {Hickey}}, \bibinfo
  {author} {\bibfnamefont {S.}~\bibnamefont {Trebst}}, \ and\ \bibinfo {author}
  {\bibfnamefont {M.~M.}\ \bibnamefont {Scherer}},\ }\href@noop {} {\bibfield
  {journal} {\bibinfo  {journal} {arXiv:1907.09490}\ } (\bibinfo {year}
  {2019})}\BibitemShut {NoStop}%
\bibitem [{\citenamefont {Natori}\ \emph {et~al.}(2019)\citenamefont {Natori},
  \citenamefont {Nutakki}, \citenamefont {Pereira},\ and\ \citenamefont
  {Andrade}}]{Natori19}%
  \BibitemOpen
  \bibfield  {author} {\bibinfo {author} {\bibfnamefont {W.}~\bibnamefont
  {Natori}}, \bibinfo {author} {\bibfnamefont {R.}~\bibnamefont {Nutakki}},
  \bibinfo {author} {\bibfnamefont {R.}~\bibnamefont {Pereira}}, \ and\
  \bibinfo {author} {\bibfnamefont {E.}~\bibnamefont {Andrade}},\ }\href@noop
  {} {\bibfield  {journal} {\bibinfo  {journal} {arXiv:1908.09224}\ } (\bibinfo
  {year} {2019})}\BibitemShut {NoStop}%
\end{thebibliography}%

\pagebreak

\setcounter{equation}{0}
\renewcommand{\theequation}{S\arabic{equation}}

\setcounter{figure}{0}
\renewcommand{\thefigure}{S\arabic{figure}}

\begin{widetext}
\begin{center}
\textbf{\large{}{}{}{}{}{}Supplementary material for ``Nematicity with a twist: rotational symmetry breaking in a moir\'e superlattice''}{\large{}{}{}{}{}{}
}
\par\end{center}

\section{Six-band tight-binding model}

Here we provide the details of implementing nematic order in the six-band tight-binding model for twisted bilayer graphene (TBG) introduced in Ref.~\onlinecite{Po2019}. We will generally follow the notation and convention of Ref.~\onlinecite{Po2019}, with the exception of a few minor modifications which simplify the notation for the present purposes. 

The six-band tight-binding model of Ref.~\onlinecite{Po2019} is defined by $(p_z,p_+,p_-)$ orbitals on the sites of a triangular lattice and $s$ orbitals on the sites of a kagome lattice, as shown in Fig.~\ref{fig1}. The purpose of the six-band model is to reproduce the low-energy flat bands of TBG in such a way that all symmetries manifestly present in the continuum description are respected (and are implemented naturally). The authors of Ref.~\onlinecite{Po2019} introduce a number of models which achieve this; here we choose the six-band model to study nematic order in TBG. 

It is important to note that the six-band model describes the two low-energy flat bands originating from \emph{a single} valley of the two graphene sheets forming the bilayer system. This implies that a model which includes the full set of flat low-energy bands (apart from spin) must have twelve bands: two copies of the six-band model related by the symmetries which exchange valleys. Here we sketch how such model is constructed and show how it can be supplemented with the appropriate symmetry breaking terms to account for nematic order.

We begin by recalling the definition of the six-band model. The orbital degrees of freedom can be represented by a fermion operator $\psi^\dagger_{\bf k} $ given by
\begin{equation}
\psi^\dagger_{\bf k} = (p^\dagger_{{\bf k}z},  p^\dagger_{{\bf k}+}, p^\dagger_{{\bf k}-}, a^\dagger_{{\bf k}}, b^\dagger_{{\bf k}}, c^\dagger_{{\bf k}}).
\end{equation}
Note that this a slight departure from Ref.~\onlinecite{Po2019}. In terms of these degrees of freedom the (six-band) Hamiltonian for a single valley is given by~\cite{Po2019}
\begin{equation}
H_{\bf k} = \begin{pmatrix} H_{p_z} + \mu_{p_z} &  C^\dagger_{p_\pm p_z} & 0 \\  C_{p_\pm p_z} & H_{p_\pm} + \mu_{p_\pm}  & C^\dagger_{\kappa p_\pm}\\ 0 & C_{\kappa p_\pm} & H_{\kappa} + \mu_{\kappa} \end{pmatrix}.   \label{eq:H6}
\end{equation}
Here $H_{p_z} $ and $H_{p_\pm} $ are the subblock Hamiltonians in the $p_z$ and $p_\pm$ subspaces, and $H_{\kappa}$ describes the coupling between the kagome lattice sites. The subblocks $C_{XY}$ describe the couplings between the $X$ and $Y$ sectors (where $X,Y=p_z,p_\pm,\kappa$). The form of all these subblocks are given in Ref.~\onlinecite{Po2019}, including the parameter set we use here (see Table VI in Ref.~\onlinecite{Po2019}).

To promote the six-band model to a full twelve-band model, we take two copies and introduce a valley degree of freedom as
\begin{equation}
\Psi^\dagger_{\bf k }=(\psi^\dagger_{\bf k +},\psi^\dagger_{\bf k -}),
\end{equation}
where $\pm$ labels the $K$ and $K'$ valleys of the individual graphene layers. The full Hamiltonian $\mathcal H_{\bf k}$ is then given by 
\begin{equation}
\mathcal H_{\bf k} = \begin{pmatrix} H_{\bf k} & \\ & U H_{-\bf k}U^\dagger \end{pmatrix}, \label{eq:H12}
\end{equation}
where $H_{\bf k}$ is the Hamiltonian of Eq.~\eqref{eq:H6} and $U\equiv U_{C_{2z}}$ is the matrix representation of the twofold rotation $C_{2z}$. A word of caution with respect to the tight-binding gauge choice is appropriate here.  Ref.~\onlinecite{Po2019} uses a gauge for which $H_{{\bf k}+ {\bf G}} = H_{\bf k}$ holds, where ${\bf G}$ is a reciprocal lattice vector. In this gauge the matrix $U$ is momentum dependent and thus takes a more complicated form. A simpler form is obtained in the more conventional tight-binding gauge, in which case matrix representations of symmetries are momentum independent. In particular, in the tight-binding gauge $U$ is given by $U=\text{Diag}(1,-1,-1,1,1,1)$. In what follows we will adhere to the gauge choice of Ref.~\onlinecite{Po2019}. 

The Hamiltonian of Eq.~\eqref{eq:H12} with the parameters specified in Ref.~\onlinecite{Po2019} defines a tight-binding model for TBG that respects all symmetries, including a $U_v(1)$ valley conservation symmetry. Various symmetry breaking terms can be considered, and here we are specifically interested in terms that break $C_{3z}$ symmetry but preserve $C_{2z}$ symmetry, as required by quadrupolar nematic order. The $U_v(1)$ valley symmetry is not directly relevant to Potts-nematic order, but it is nonetheless an important property of the TBG system and therefore it is useful to specify whether or not it is preserved by additional symmetry breaking terms. 

Let us first note that a coupling of the valleys of the form
\begin{equation}
\delta \mathcal H  = \Delta \sum_{\bf k} \psi^\dagger_{\bf k +}\psi_{\bf k -}+\text{H.c.},  \label{eq:valley-break}
\end{equation}
breaks the $U_v(1)$ valley but respects all lattice symmetries as well as time-reversal symmetry. Added to the Hamiltonian of Eq.~\eqref{eq:H12} it enters as an off-diagonal block. The Fermi surfaces of the Hamiltonian in the absence and presence of Eq.~\eqref{eq:valley-break} are shown in Figs.~\ref{fig2}(a) and (b), respectively.

\begin{figure}
\includegraphics[width=0.45\columnwidth]{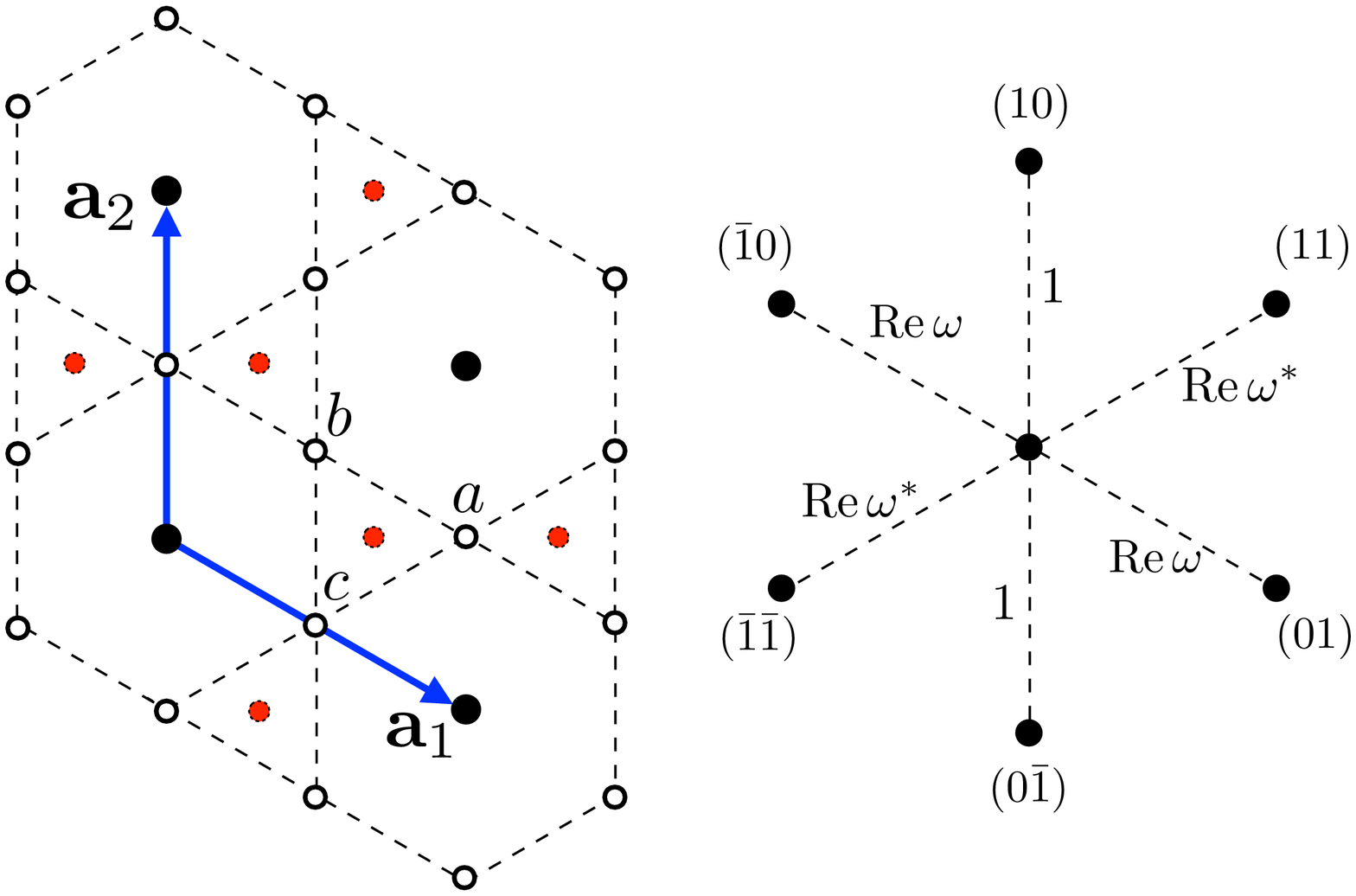} \caption{\label{fig1} The model of Ref.~\onlinecite{Po2019} is defined by $(p_z,p_+,p_-)$ orbitals on sites of a triangular lattice (indicated by black solid dots) and $s$ orbitals on the sites of a kagome lattice (indicated by solid white dots). Note that the kagome lattice sites are located on the edges of the triangular Wigner-Seitz cell. The Bravais lattice vectors ${\bf a}_{1,2}$ used in Eq.~\eqref{eq:phases} are shown in blue. The three kagome sites in the unit cell are labeled $(a,b,c)$. The position of the honeycomb lattice sites are indicated by red dots, but do not play a role in our analysis.}
\end{figure}

\section{Rotational symmetry breaking}

Next, we consider the rotational symmetry breaking terms that constitute nematic order. Since the model is built from multiple degrees of freedom, there are a number of different ways in which rotation symmetry breaking can be implemented. We first focus on the triangular lattice sector of the model. Within this sector there are two possibilities: nematic order can occur as a result of hopping anisotropy or due to a lifting of the orbital degeneracy. To model the first possibility we introduce the two $d$-wave form factors
\begin{eqnarray}
d_{{\bf k} 1} & = & \phi_{01} + \text{Re}\,\omega^*\, \phi_{\bar 1 \bar 1} + \text{Re}\,\omega \;\phi_{10},  \label{eq:d1} + \text{c.c.} \\
d_{{\bf k} 2} &= & \text{Im}\,\omega^*\, \phi_{\bar 1 \bar 1} + \text{Im}\,\omega \;\phi_{10} + \text{c.c.},   \label{eq:d2}
\end{eqnarray}
where the phases $\phi_{lm}$ are defined as (see Ref.~\onlinecite{Po2019})
\begin{equation}
\phi_{lm} = e^{-i {\bf k}\cdot (l {\bf a}_1 + m {\bf a}_2)}. \label{eq:phases}
\end{equation}
Here we use the notation $\bar l \equiv -l$ and $
\omega= \exp(2\pi i/3)$. The hopping pattern of the $d$-wave form factor $d_{{\bf k} 1} $ is graphically shown in Fig.~\ref{fig1}. These form factors have precisely the same symmetry as $(\Phi_1,\Phi_2)$ introduced in the main text. 

The triangular lattice $d$-wave form factors can then be used to introduce a symmetry breaking perturbation in the triangular lattice ($p$-orbital) sector. For instance, we can add a perturbation $\delta H_{p_z} $ to the Hamiltonian $H_{p_z}$ of the $p_z$ orbital appearing in Eq.~\eqref{eq:H6} given by
\begin{equation}
\delta H_{p_z}  = \Phi_1 d_{{\bf k} 1}+\Phi_2 d_{{\bf k} 2}, \label{eq:dHpz}
\end{equation}
where $\Phi_{1,2}$ are the nematic order parameters as defined in the main text. This term gives the Fermi surface distortion of Fig. 4(b) of the main text. Clearly, the same perturbation (but proportional to the appropriate identity matrix) can be added to $H_{p_\pm}$, which describes the $p_\pm$ orbitals.

\begin{figure}
\includegraphics[width=0.4\columnwidth]{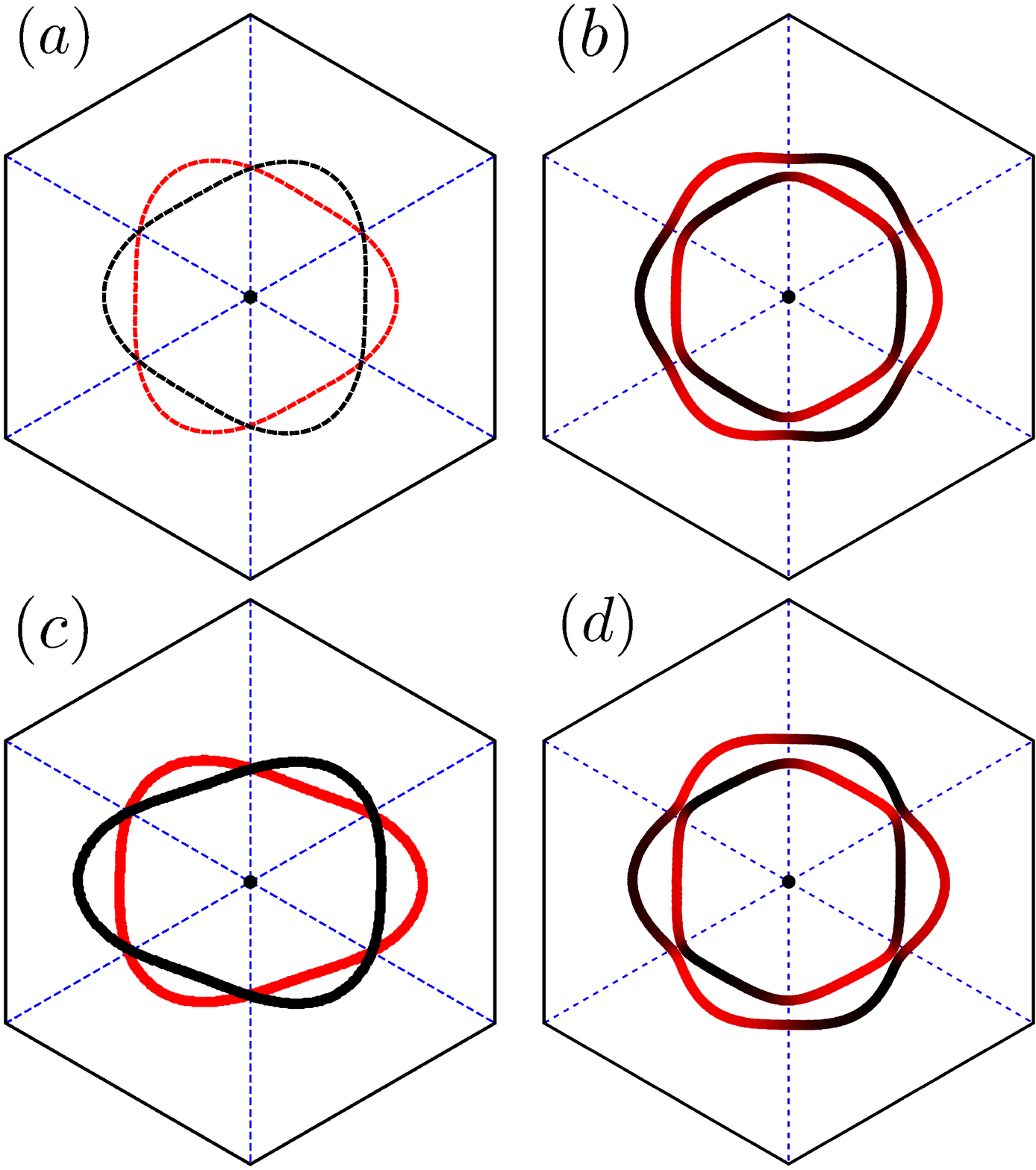} \caption{\label{fig2} Fermi surface of the 6-band model of Ref.~\onlinecite{Po2019} with parameters as specified in Ref.~\onlinecite{Po2019} ; red and black correspond to the two valleys of the individual graphene sheets (reproduced from the main text). (b) Fermi surface of the same model as in (a) but with the additional $U_v(1)$ valley symmetry breaking terms given by Eq.~\eqref{eq:valley-break}. Colors (black and red) correspond to the projection of eigenstates on the two valleys. (c) Fermi surface in the presence of (intra-valley) orbital nematic order given by Eq.~\eqref{eq:dHpxy}. (d) Fermi surface in the presence of \emph{inter}-valley nematic order given by Eq.~\eqref{eq:dH-valley}. }
\end{figure}

In the $p_\pm$-orbital sector the nematic order parameter couples to another symmetry breaking perturbation, which is independent of momentum. Making the two orbitals inequivalent lifts their degeneracy and necessarily breaks threefold rotation symmetry. In particular, the perturbation $\delta H_{p_\pm} $ which achieves this is given by
\begin{equation}
\delta H_{p_\pm}  = \begin{pmatrix} 0 &  \Phi_1 -i \Phi_2  \\   \Phi_1 + i \Phi_2& 0 \end{pmatrix}.  \label{eq:dHpxy}
\end{equation}
Note that the diagonal terms are zero since time-reversal symmetry must be preserved. (An overall energy can be absorbed in $\mu_{p_\pm}$.) The Fermi surface in the presence of a nematic distortion given by $\delta H_{p_\pm} $ is shown in Fig.~\ref{fig2}(c). As argued and expected, the distortion is qualitatively similar to a distortion originating from $d$-wave form factors in the kinetic terms (Fig. 4(b) of the main text).

In Fig.~\ref{fig3}, we show the change in the Fermi momentum due to nematic order, $\delta k_F$, corresponding to the Fermi surface of Fig. 4(b) of the main text. Blue and yellow denote positive and negative values, respectively. First, we see that the shape of $\delta k_F$ corresponds to what one expects from the $d$-wave form factor in hexagonal lattices. Second, we note that the maximum $\delta k_F$ occurs precisely at the hot spot identified in Fig. 4(a) of the main text.

Consider next the kagome lattice sector of the model. The kagome sector does not have an orbital degree of freedom but it does have multiple sites in the unit cell. The simplest coupling to the nematic order parameter is given by a charge ordering perturbation within the unit cell, which breaks threefold rotations but preserves the twofold rotation $C_{2z}$. Specifically, the perturbation $\delta H_{\kappa}$ to the kagome lattice Hamiltonian $H_{\kappa}$ is given by
\begin{equation}
\delta H_{\kappa}= ( \Phi_1 - i \Phi_2 ) \begin{pmatrix} 1 & & \\ & \omega  &  \\ & & \omega^* \end{pmatrix} + \text{H.c.},   \label{eq:dHk}
\end{equation}

\begin{figure}
\includegraphics[width=0.45\columnwidth]{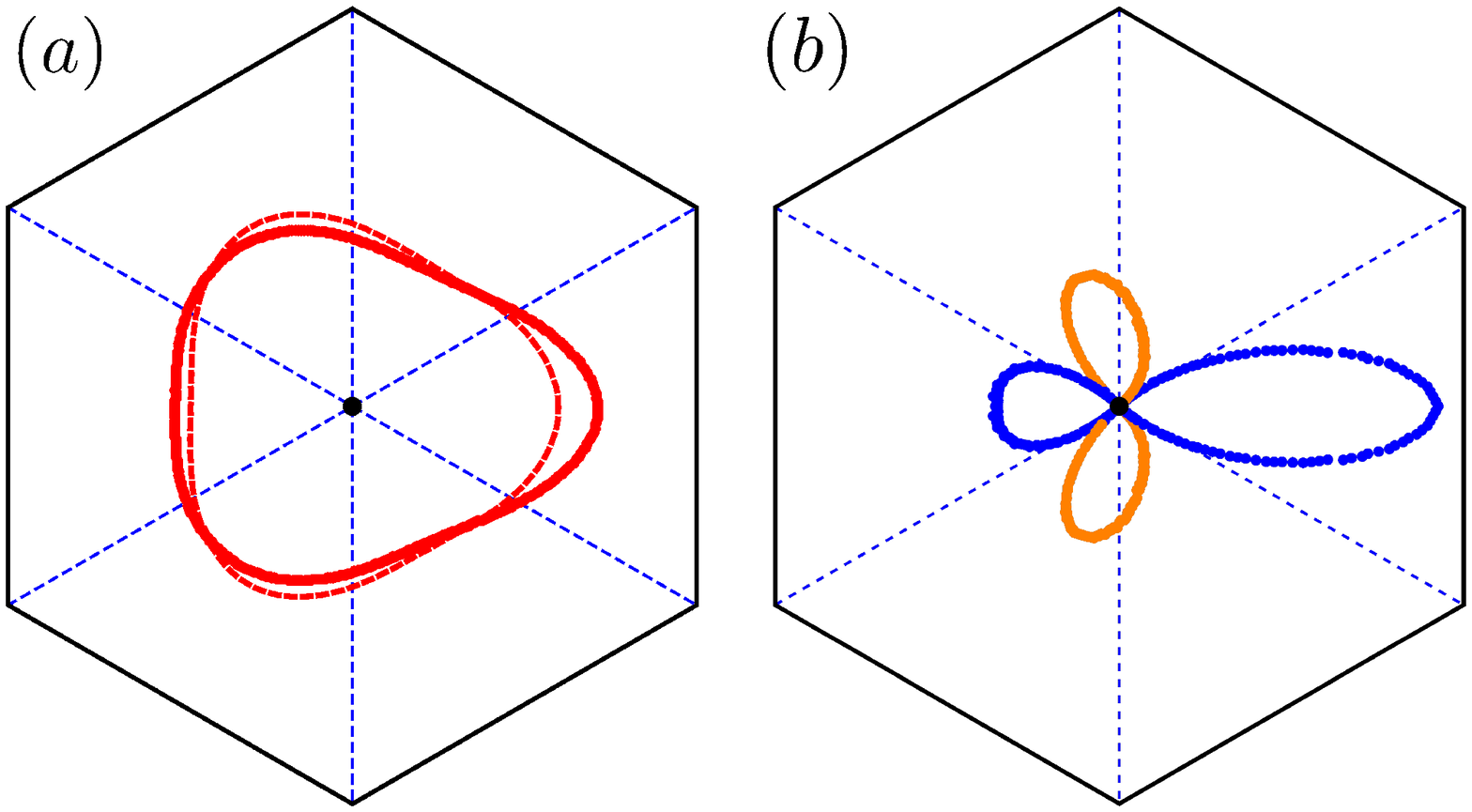} \caption{\label{fig3} To demonstrate the $d$-wave symmetry of the Fermi surface distortion in the presence of intra-valley nematic order we show $\delta k_F$, i.e. the change in Fermi momentum, as function of angle. Panel (a) shows the (single-valley) distorted Fermi surface of Fig. 4(b) of the main text and panel (b) shows $\delta k_F$, which clearly exhibits four nodes and a sign change (indicated by different colors) at the nodes.}
\end{figure}

The rotational symmetry breaking perturbations introduced so far are all intra-valley perturbations; they should be considered as perturbations to Eq.~\eqref{eq:H6}, with the full Hamiltonian given by the prescription of Eq.~\eqref{eq:H12}. One may, however, also consider inter-valley nematic coupling terms, which enter the off-diagonal blocks in Eq.~\eqref{eq:H12}. More precisely, the Hamiltonian of Eq.~\eqref{eq:H12} is modified according to 
\begin{equation}
\mathcal H_{\bf k} \rightarrow \begin{pmatrix} H_{\bf k} & \\ & U H_{-\bf k}U^\dagger \end{pmatrix} + \delta \mathcal H_{\Phi}, \label{eq:H12-mod}
\end{equation}
where $\delta \mathcal H_{\Phi}$ collects all terms which describe nematic distortions and takes the form
\begin{equation}
\delta \mathcal H_{\Phi}=  \begin{pmatrix}  & \Delta_{\Phi} \\  \Delta^\dagger_{\Phi} &  \end{pmatrix}. \label{eq:dH12}
\end{equation}
The form of $\Delta_{\Phi} $ depends on the choice of nematic coupling; as in the case of intra-valley nematic coupling, in principle many possibilities of inter-valley nematic coupling exist. One simple type of nematic coupling is given by
\begin{equation}
\Delta_{\Phi} = \delta H_{p_z} \oplus \delta H_{p_\pm}  , \label{eq:dH-valley}
\end{equation}
where $\delta H_{p_z} $ and $\delta H_{p_\pm} $ are given by Eqs.~\eqref{eq:dHpz} and~\eqref{eq:dHpxy}. The Fermi surface corresponding to inter-valley nematic coupling of this form is shown in Fig.~\ref{fig2}(d).

We have based our microscopic discussion of rotation symmetry breaking on the model introduced in Ref.~\onlinecite{Po2019}. This was motivated by the natural implementation of all relevant symmetries in this model. It is worth stressing that an analysis of nematic order in TBG similar to the one presented here can also be obtained from different microscopic (tight-binding) models proposed for TBG~\cite{Yuan2018,Po2018,Koshino2018,Zou2018,Kang2019}. 

\end{widetext}

\end{document}